\pdfoutput=1 
\documentclass{nature}

\usepackage{multirow}
\usepackage{ccaption}
\usepackage{subcaption}
\usepackage{graphicx}
\usepackage{lipsum}
\usepackage{ragged2e}
\usepackage{fixltx2e}
\usepackage{hyperref}

\usepackage{comment}
\usepackage{cite}
\usepackage{adjustbox} 

\makeatletter
\newenvironment{figurehere}
{\def\@captype{figure}}
{}
\makeatletter

\usepackage{xcolor}

\title{\begin{center}  Hardware Functional Obfuscation With Ferroelectric Active Interconnects \end{center}}

\author
{ \centering
{
\large{Tonggunag Yu$^{1}$, Yixin Xu$^{1}$, Shan Deng$^{2}$, Zijian Zhao$^{2}$, Nicolas Jao$^{1}$, \\
You Sung Kim$^{3}$, Stefan Duenkel$^{3}$, Sven Beyer$^{3}$, Kai Ni$^{2*}$, Sumitha George$^{4*}$, Vijaykrishnan Narayanan$^{1}$\\}
\vspace{3ex}
\normalsize{$^{1}$Pennsylvania State University, State College, PA 16802, USA}\\
\normalsize{$^{2}$Rochester Institute of Technology, Rochester, NY 14623, USA}\\
\normalsize{$^{3}$GLOBALFOUNDRIES Fab1 LLC \& Co. KG, Dresden, Germany}\\
\normalsize{$^{4}$North Dakota State University, Fargo, ND 58102, USA}\\
\vspace{2ex}
\normalsize{$^{*}$To whom correspondence should be addressed} \\
\normalsize{Email: sumitha.george@ndsu.edu, kai.ni@rit.edu} \\
}}

\begin{document}
\flushbottom
\maketitle
\begin{abstract}
Camouflaging gate techniques are typically used in hardware security to prevent reverse engineering. Layout level camouflaging by adding dummy contacts ensures some level of protection against extracting the correct netlist. Threshold voltage manipulation for multi-functional logic with identical layouts has also been introduced for functional obfuscation. All these techniques are implemented at the expense of circuit-complexity and with significant area, energy, and delay penalty. In this paper, we propose an efficient hardware encryption technique with minimal complexity and overheads based on   ferroelectric field-effect transistor (FeFET) active interconnects.
The active interconnect provides run-time reconfigurable inverter-buffer logic by utilizing the threshold voltage programmability of the FeFETs.  Our method utilizes only two FeFETs and an inverter to realize the masking function compared to  recent reconfigurable logic gate  implementations using several FeFETs and complex differential logic. We fabricate the proposed circuit and demonstrate the functionality. 
Judicious placement of the proposed logic in the IC makes it acts as a hardware encryption key and  enables encoding and decoding of the functional output without affecting the critical path timing delay. Also, we achieve comparable encryption probability with a limited number of encryption units. In addition, we show a peripheral programming scheme for reconfigurable logic by reusing  the existing scan chain logic, hence obviating the need for specialized programming  logic and circuitry for keybit distribution.
Our analysis shows  an average encryption probability of 97.43\% 
with an increase of 
2.24\%/ 3.67\% delay for the most critical path/ sum of 100 critical paths delay for ISCAS85 benchmarks.  
\end{abstract}

%
%
\thispagestyle{empty}

\section*{\large Introduction}



Hardware security is becoming increasingly prominent with globalization and outsourcing integrated circuit (IC) fabrication to various foundries\cite{LogicLocking}.
Recent incidents prove that reverse engineering (RE) is possible even in advanced technology nodes. For example, there are reports of revealing the details of  processors of Apple iPhone5\cite{Apple,Mai'16}, extracting information of the baseband processor from Texas Instruments\cite{StateOfArt}, analyzing manufacturing technology of Intel's CPU\cite{intel} etc, through reverse engineering. Objects ranging from large aircraft to the smallest microchips are vulnerable to reverse engineering\cite{Rahman'14}. Attackers' motives may include commercial piracy, intelligence, patent laws\cite{StateOfArt}. RE techniques can enable the attacker to inject a hardware Trojan, copy propriety IPs, extract hard-coded keys, and copy instruction sequences\cite{GDStoNetlist}. Such scenarios necessitate the need for hardware encryption in chips, which adds a level of difficulty to IC analysis\cite{Randy'09} and reverse engineering. 

Reverse engineering extracts information from an integrated circuit utilizing techniques like depackaging, delayering, high resolution imaging and side-channel probing, etc\cite{StateOfArt}. 
For example,  attackers often depackage the target chip, take high definition image of each layer, and then use an image recognition software to extract netlists\cite{GDStoNetlist}. Different layout shapes of different logic cells make this process easy for the attackers to gather logic information. To mitigate such risks, an effective technique is to add camouflaged cells in the design such that discerning logic process through reverse engineering is difficult or impossible. Camouflaged cells prevent the interpretation of correct functionality by being logically obscured. 

Various gate-level camouflaging techniques have been developed with conventional CMOS devices\cite{Mai'16, ISSCC'18, ETS'16, HOST'18, ISLPED'18}. Traditional CMOS-based camouflaging implementations incur overheads in circuit area, power, and delay. Recent explorations have investigated emerging devices 
such as  spin-transfer-torque devices\cite{VLSI'14}, tunnel-FETs \cite{ESTCS'14}, ferroelectric devices \cite{ESTCS'15,2D-polarity}, tungsten diselenide (WSe\textsubscript{2}) devices\cite{nature'20}, etc., for provisioning hardware security by leveraging unique properties such as their non-volatile behavior. 
Rajendran et al.\cite{CCS'13} proposed a gate camouflaging technique by inserting dummy via/contacts in the layout and creating look-alike layouts for  NAND, NOR, and XNOR cells (Fig.~\ref{fig:overview}(c)).
With layout look-alike camouflaged gates, the attackers may interpret the function incorrectly and end up with a faulty netlist.
However, advances in imaging and computer vision technology have made such methods less effective and  susceptible to direct probing attacks\cite{Mai'16}. One approach to avoid probing attacks is to use internal parameters of devices, such as product invariability or different states of the devices, for the implementation of different functions while retaining identical physical layouts\cite{Mai'16,S_Datta_reconfig}.  
 Wu et al.\cite{2D-polarity} proposed that two-dimensional black phosphorus field-effect transistors with reconfigurable polarities are suitable for hardware security applications (Fig.~\ref{fig:overview}(c)). These transistors can be dynamically switched between p-FET and n-FET operations through electrostatic gating. Though this approach achieves minimum area overhead, its integration with Si CMOS technology is challenging. 

Erbagci et al.\cite{Mai'16} proposed a  gate camouflaging technique using threshold voltage defined (TVD) logic topology. The key idea relies on the usage of different threshold voltage (\textit{V}\textsubscript{TH}) transistors but with identical physical layouts. Their work introduced a  generic 2-input TVD logic gate capable of realizing multiple logic functions (NAND, NOR, and XNOR). They achieved this by setting pull-down transistors with  different \textit{V}\textsubscript{TH} implantations ( i.e. low-\textit{V}\textsubscript{TH} (LVT), and high-\textit{V}\textsubscript{TH} (HVT)). However, this circuit does not provide flexible reconfigurability as the \textit{V}\textsubscript{TH} of conventional CMOS transistors are not run-time programmable. 
Dutta et al.\cite{S_Datta_reconfig} further enhanced the TVD device design (Fig.~\ref{fig:overview}(c)) by replacing the pull-down logic transistors with emerging ferroelectric FETs (FeFETs). By utilizing the feature of voltage-dependent polarization switching of FeFET, pull down transistors can be reprogrammed into LVT and HVT states. Exploiting the programmable \textit{V}\textsubscript{TH} of the FeFETs makes the TVD logic gate-level camouflaging and run-time reconfigurable simultaneously. However, these features come at the expense of complex design with differential logic, high area, power, timing expense. 

\begin{figurehere}
   \centering
    \includegraphics[scale=1,width=\textwidth]{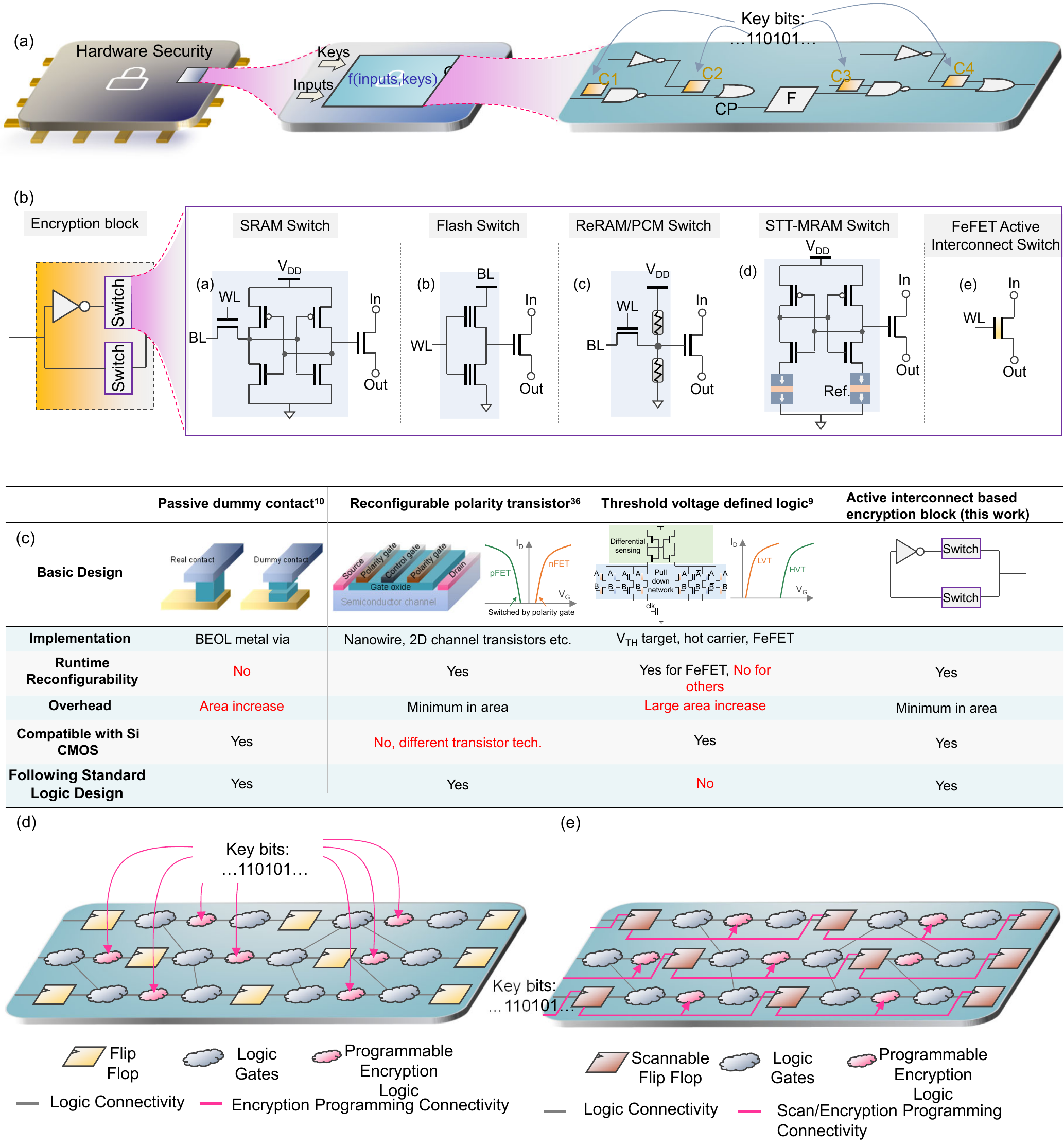}
    \captionsetup{parbox=none} 
    \caption{Overview of the proposed encryption of  IC logic design harnessing the ultra-compact FeFET active interconnect reconfigurable switch. (a) Illustration of utilizing the active interconnect based encryption block for obfuscating the IC logic.  The keys are to dynamically reconfigure the interconnect such that the logic function is hidden. (b) The encryption  block and implementations of  switches using various technologies to realize  camouflaging logic function. FeFET realizes a reconfigurable ultra-compact  active interconnect switch. (c) Comparison between different camouflage logic designs. The proposed active interconnect based approach is advantageous in realizing the camouflage logic. (d) Distribution challenge of key bits to the programmable encryption logic block. (e)  Distribution of key bits to the programmable encryption logic block via scan chain logic. }
    \label{fig:overview}
\end{figurehere}

This work proposes a simple area efficient reconfigurable logic which can act as encryption key logic. To avoid IC counterfeiting,  the functional IP is locked  with key logic and the IP  can be unlocked with a correct sequence of keys given to the trusted customer. Therefore protection can be achieved by intentionally programming the device with incorrect keys.
 Our proposed scheme for securing the ICs by hardware encryption is shown in Fig.~\ref{fig:overview}(a). In this scheme, an active interconnect based encryption block is designed  and is chosen to judiciously place them at different locations in the chip. Keybits are used to program encryption logic.  
 An example arrangement of encryption blocks C1,C2,C3 \& C4 in an IC is shown in Fig.~\ref{fig:overview}(a). In this scheme, not all gates need to be camouflaged. 
Rajendran et al.\cite{CCS'13} has shown that choosing a subset of gates to be  camouflaged is sufficient to make the IC immune to reverse engineering. 
One advantage with this simple yet but powerful technique is that, 
the placement of key logic can be in the {non-critical} timing branch of the logic, yet the output function will be encrypted.  Further analysis shows that the strategic placement of the key circuit influences the output without posing challenges in the timing closure. In this way, the proposed scheme causes only minimal interference to the actual circuit in terms of delay, area and power.

It is crucial to develop an encryption logic with efficient functional implementation, resistance to  hardware attacks, CMOS compatibility, high density and minimal overheads. The ability to program the encryption block multiple times  during run-time is significant for enhanced security.  To satisfy aforementioned requirements,  a compact encryption block is designed as shown in Fig.~\ref{fig:overview}(b). Here  the block needs to be developed in such a way that  the output signal gets inverted or non-inverted  based on the internal state of the switch. The switch can be either in closed or in open state. Note that complementary states are maintained in the upper and lower branches.  If the switch in the upper branch is in a  conducting (closed) state then inverted input appears in the output and if the switch in the lower branch is in a conducting (closed) state then the output follows input. Thus with reconfigurability, the proposed encryption block can act as either an inverter or buffer.  The vision is to integrate the reconfigurability to  switch's internal state such that  physical layout looks identical in both modes of operation. Hence  the block can act as a camouflaging buffer-inverter standalone gate. In addition,  the usage of this block in conjunction  with other complex gates by placing it at their input or output, the overall functionality changes, extending the camouflage to complex circuits. 

There are many promising memory technologies available to realize the switch in the reconfigurable encryption block, each with its own unique features. Fig.~\ref{fig:overview}(b) shows the potential implementations of the switch using SRAM, Flash, resistive RAM (ReRAM), phase change memory (PCM), spin transfer torque magnetic RAM (STT-MRAM), and ferroelectric field-effect  transistors (FeFETs).
SRAM is the most straightforward memory to use but is volatile and typically requires at least six transistors, dissipating significant leakage power while suffering from low memory density. A Flash memory-based switch is nonvolatile and compact \cite{greene201165nm}, but memory programming is slow ($\sim$ms) and requires a high programming voltage ($\sim$10 volts). Emerging nonvolatile memory (NVM)-based switches have also been proposed and superior performance
has been demonstrated \cite{huang2014low, zhao2009spin, tanachutiwat2010fpga}. Resistive memories are a class of two terminal NVM devices, including ReRAM, PCM, and STT-MRAM. Information is stored as conductive filament formation or rupture (ReRAM), film crystallization or amorphization (PCM), or parallel or anti-parallel orientation of the magnetization in a magnetic tunnel junction (STT-MRAM). These devices are nonvolatile and compact, but usually require a large conduction current to program the devices, consuming a significant write power. The limited on/off resistance ratio ($\sim$100 for ReRAM/PCM and $\sim$5 for STT-MRAM) usually requires additional circuitry, such as the 1T2R structure in ReRAM/PCM \cite{huang2014low, tanachutiwat2010fpga} and an even more complex supporting structure for STT-MRAM \cite{zhao2009spin} to realize a nonvolatile switch.

In this work, a nonvolatile active interconnect switch,  based on a single FeFET is proposed to build the reconfigurable encryption block. In a FeFET, the ferroelectric layer is integrated as the gate dielectric of a MOSFET, where the information is stored in the direction of the ferroelectric polarization, which can be switched with an applied electric field. By configuring the direction of the polarization to point towards the semiconductor channel or the gate electrode, the device is set to either low-\textit{V}\textsubscript{TH} or high-\textit{V}\textsubscript{TH} state respectively. This makes FeFET an integrated single transistor memory, a great advantage to realize the nonvolatile switch. The dynamic reconfigurability of \textit{V}\textsubscript{TH} state has been harnessed in many applications, for instance, on memory-centric computing \cite{Ma'16, Xueqing'18, Lai'18, ni2019ferro,8352114, S_Datta_reconfig, 9073266}.
Since the ferroelectric memory is written with an electric field rather than a large conduction current \cite{ni2019ferro}, this technology becomes highly energy efficient (e.g., down to $\sim$1 fJ/bit write energy). Therefore the dynamic reconfigurability of \textit{V}\textsubscript{TH}, along with its intrinsic three-terminal structure, nonvolatility, superior write performance, excellent CMOS compatibility, and scalability \cite{khan2020future,mikolajick2020past}, shape FeFET a prime candidate for the nonvolatile switch. 
The three terminal device structure makes the FeFET a very compact active interconnect.
The key idea behind our proposed active interconnect based reconfigurable encoding circuit leverages the run time reconfigurability of the encryption block by manipulating the threshold voltage of FeFET \cite{IEDM'16, IEDM'17, IEDM'181, Jerry_2018, Kai'18}.  


Note that design variants of our proposed active interconnect based  dynamically configurable  block can be extended to offer various chip design applications. For example, an active configurable route switching can be enabled, as shown in Fig. \ref{fig:other_applications}, to route a signal to different functional units. The directions can be tuned with programming configuration. Another example is a configurable path connector that connects/disconnects inputs to destination units. This is especially beneficial for controlling the logic towards redundant functional units. Redundant functional units are typically used in chips as means to increase the reliability against fault tolerance. Reconfigurable logic gate is another byproduct of the proposed method which is realizable by programming the control inputs gates. Many combinations  such as inverter, NAND, AND, OR, NOR, XOR and XNOR are possible (Fig.~\ref{fig:other_applications}). In addition, recongfigurable gates can be deployed in chips to tackle Engineering Change Orders (ECO)\cite{5325880,4359938}, where  functional logic changes need to be met with minimal layout changes. The ability to meet functional changes with existing gates in the design is relevant for both pre-mask and post-mask ECOs.

All dynamic logic programming schemes  including the aforementioned dynamic encryption programming pose a challenge in getting the desired  input values to the configurable logic which mostly requires  appropriate write voltages to set its internal state. This necessitates a robust peripheral logic and circuitry. However, a systematic approach for peripheral programming has rarely been explored in recent reconfigurable logic research works. Nowadays application specific integrated circuit(ASIC)/system on a chip (SoC) implementations come with more than a million gates and flipflops spread across the entire chip. As the amount of logic increases, the number of encryption gates is also expected to increase proportionally. 
In such cases, it is not trivial to program umpteen configurable gates.
Fig.~\ref{fig:overview}(d) shows an example distribution of logical blocks that need to be programmed in a dynamically configurable security circuit. Explicit addition of  auxiliary logic and peripheral circuitry is required to support dynamic programming of the configurable logic in this case. The amount of additional logic  required  and the resultant overhead, increase with the amount of programmability incorporated in the chip,  which does not favor turning  all gates reconfigurable.

In order to eliminate the dedicated auxiliary distribution logic requirements,  this work proposes to integrate dynamic programmable encryption key distribution with the  existing scan logic in the chip. In a typical functional unit implementation, logic gates are placed between flipflops, which are clocked at a designated frequency. In many systems including IBM POWER/SYSTEM Z\cite{ibmz}, flipflops are designed to  handle both logic and scan data along with  data clocking and scan clocking.
The proposed  key distribution solution is shown in Fig.~\ref{fig:overview}(e), where scan flipflops provide encryption keybits to program the reconfigurable logic.
 Temporal sharing of  the resources is possible since the scan programming and keybit programming do not overlap in time. Reuse  and temporal sharing of the existing scan resources obviate   the need for  additional complex  logic programming and circuits, eliminate  numerous multiplexer units required  in a  specialized dynamic input  distribution unit,  etc., thus leading to minimal perturbation in the original chip.

The rest of the article is organized as follows.    Experimental verification of our proposed  reconfigurable encryption block is discussed first.
Details on SPICE simulation for functional verification of the circuit is followed.  Our placement strategy and  analysis of encryption probability on ISCAS benchmarks are discussed in subsequent sections. Peripheral programming strategies for reconfigurable blocks are discussed further. The results and conclusions are discussed in the final section. 
In addition, a supplementary section with additional  details on device  fabrication, circuit simulation, variation analysis, placement impact on encryption probability, and,  block layout and analysis is also included.

\section*{\large Verification of the Reconfigurable Block}

To verify the functionality of the proposed reconfigurable block, measurements and circuit simulations are performed. For experimental demonstration, industrial 28nm high-k metal gate FeFET devices are tested, as shown in the transmission electron microscopy (TEM) cross-section images of the device Fig.~\ref{fig:figure2_fefet}(a)\cite{trentzsch201628nm}. The device features an 8 nm thick doped HfO\textsubscript{2} as the ferroelectric layer and around 1nm SiO\textsubscript{2} as the interlayer in the gate stack, as shown in Fig.~\ref{fig:figure2_fefet}(b). The FeFET memory performance is characterized by standard \textit{I}\textsubscript{D}-\textit{V}\textsubscript{G} measurements after applying $\pm$4 V, 1$\mu$s write pulses on the gate. Fig.~\ref{fig:figure2_fefet}(c) shows a memory window about 1.2 V, i.e., the \textit{V}\textsubscript{TH} separation between the LVT and HVT states, which enables a large ON/OFF conductance ratio. Note that, for the nucleation-limited polarization switching \cite{mulaosmanovic2017switching,mulaosmanovic2020investigation}, a tradeoff can be realized between the write pulse amplitude and pulse width, as shown in Fig.~\ref{fig:figure2_fefet}(d), which presents the switching dynamics of FeFET as a function of applied pulse width for different pulse amplitudes. It clearly suggests that 4 V is not absolutely necessary and lower write voltages possible with a tradeoff of a large pulse width. This could help alleviate the design of peripheral supporting circuitry with lower write voltages in applications where the FeFET configuration is occasional and high speed write operation is not necessary, as the proposed logic camouflaging application in this work.


\begin{figurehere}
   \centering
    \includegraphics[scale=1,width=\textwidth]{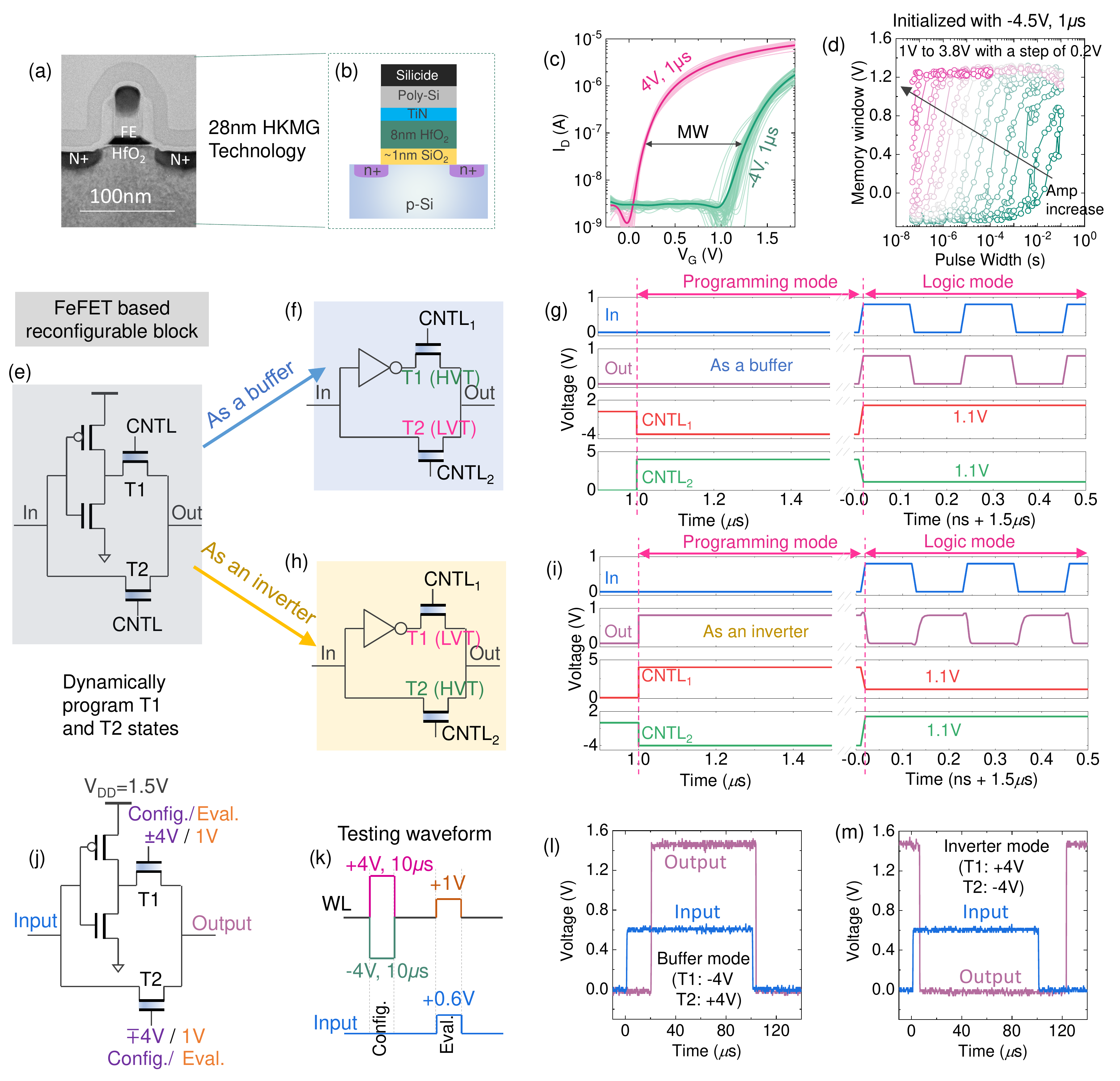}
    \captionsetup{parbox=none}
    \caption{FeFET reconfigurable encryption block and its functionality verification. (a) TEM cross section and (b) schematic cross section of a FeFET integrated on 28nm HKMG industrial technology platform. (c) \textit{I}\textsubscript{D}-\textit{V}\textsubscript{G} characteristics of 60 different FeFETs measured after $\pm$4 V, 1 $\mu$s write pulses. Good control over the device variability and a memory window of 1.2 V are demonstrated. (d) The dynamic switching characteristics of the FeFET as a function of write pulse width at different pulse amplitudes. Tradeoff between amplitudes and pulse widths are present. (e) Schematic of the proposed active interconnect based reconfigurable encryption block. (f) Buffer mode configuration and (g) Simulated waveforms in buffer mode showing the programming and logic modes (h) Inverter mode configuration (i) Simulated waveforms in inverter mode.  (j) Applied voltages on the encryption block in experiment. (k) The applied waveform for functionaltiy verification in experiment. Captured transient waveforms in the logic evaluation mode for (l) buffer and (m) inverter modes. For all tested FeFETs, $W$/$L$=500 nm/500 nm are used.}
    \label{fig:figure2_fefet}
\end{figurehere}

As mentioned above, the large memory window and ON/OFF conductance ratio present a unique opportunity for FeFETs to design an active interconnect based camouflaging pass transistor (switch). In addition, the capability to dynamically shift the \textit{V}\textsubscript{TH} makes the proposed active interconnect based FeFET-switch  immune  for the attacker to reverse engineer the netlist simply from layout (GDS) level. 
Fig.~\ref{fig:figure2_fefet}(e) shows the implementation of the proposed encryption block utilizing \textit{V}\textsubscript{TH} manipulation. The proposed encryption block consists of an inverter and two FeFETs.
It operates in two modes, the programming mode ,and the logic mode.  In programming mode, relatively high write voltages are used to program the device to set the \textit{V}\textsubscript{TH}. Once it is programmed, the device is all set to operate in the logic mode, where a small read voltage at the gate between the \textit{V}\textsubscript{TH} of LVT and HVT states is applied to read the FeFET state. The pass transistor either conducts or blocks the input signal based on the internal programmed \textit{V}\textsubscript{TH} state, as shown in Fig.~\ref{fig:figure2_fefet}(f) and (h), respectively. Hence  an inverted input or a non inverted input is obtained at the output of our proposed reconfigurable circuit.

The functionality of the proposed reconfigurable buffer-inverter encryption block has been verified in SPICE simulations, using a calibrated FeFET model \cite{deng2020comprehensive} and 45 nm ( NCSU FreePDK) logic transistor technology \cite{freepdk45}. 
 In the program mode,  \textit{V}\textsubscript{TH} of the two FeFETs pass transistors are set by applying write pulses. For this study, write pulses of $\pm$4 V, 500 ns are adopted. In the logic mode, an INPUT signal of 0.8 V at the \textit{In} terminal and control signal of 1.1 V(read voltage chosen between the \textit{V}\textsubscript{TH} of LVT and HVT states) at the gates of FeFETs (\textit{CNTL1} and \textit{CNTL2}) is asserted. In the buffer mode of encryption, as illustrated in Fig.~\ref{fig:figure2_fefet}(f), FeFETs T1/T2 are written into HVT/LVT state respectively by asserting  write voltages in  CNTL1 and CNTL2 terminals, as shown in the transient waveform in Fig.~\ref{fig:figure2_fefet}(g). In the logic (evaluation) mode,  it can be seen that the output( \textit{Out}) follows the input (\textit{In}) as shown in Fig.~\ref{fig:figure2_fefet}(g). On the other hand, by writing the FeFETs T1/T2 into LVT/HVT state, the inverter mode of encryption shown in Fig.~\ref{fig:figure2_fefet}(h) can be realized, which will output an inverted input signal during the logic evaluation mode, as shown in Fig.~\ref{fig:figure2_fefet}(i).

The reconfigurability of the inverter-buffer block has also been verified experimentally using the testing setup shown in Fig.~\ref{fig:figure2_fefet}(j) and (k). Discrete inverter and FeFETs are assembled together for experimental verification. The relevant applied voltages are shown in Fig.~\ref{fig:figure2_fefet}(j) and (k). The buffer mode and inverter mode operations are shown in Fig.~\ref{fig:figure2_fefet}(l) and (m), respectively. Here only the evaluation phase waveforms are shown for clarity. Correct operations of both working modes are demonstrated. Due to the large parasitics present in the testing setup, the speed is limited to tens of $\mu$s. But it is expected that with the fully integrated circuit, high speed operations can be achieved as demonstrated in the SPICE simulations in Fig.~\ref{fig:figure2_fefet}(g) and (i). In summary, the proposed encryption circuit with the same input results in two different logical outputs based on the programmed states of FeFETs, making it a strong candidate for reverse engineering resilient hardware (Fig.~\ref{fig:figure2_fefet}(g),(i)). In the next section, the encryption probability and timing beneficial placement  are discussed.

\section*{\large Encryption \& Criticality of Placement } \label{sec:kk}

For the proposed active interconnect based encryption blocks to function and enable resistance for reverse engineering, a judicious placement of the blocks enabling the encryption engine in a timing aware fashion is required. 
The location of the placement, the neighboring cells and input pattern have an impact on encryption. All these factors can contribute to logic masking effect and prevent the encrypted bit propagation to the output. In order to  understand the impact of an encrypted bit and how it propagates,  a circuit having a single encryption bit cell is analysed first.  Fig.~\ref{fig:figure4_propagation}(a) shows an example of an encrypted circuit. C1 is the proposed buffer-inverter encryption block.
Suppose I$_1$, I$_2$, I$_3$, I$_4$ to be 0, 0, 0, 1. With C1 programmed in buffer mode, the output will be  a "0".  However C1 in inversion mode will alter the output to be a "1". Note, if the input I$_2$ changes to bit "1" as shown in Fig.~\ref{fig:figure4_propagation}(b), the inverted bit from C1 will not make an impact on the output, as OR gate with input 1 masks the other input. To improve the encryption strength, additional encryption key circuit can be inserted as shown in Fig.~\ref{fig:figure4_propagation}(c).
Note, 100 \% inversion in the output all the time is also not reliable for security purposes as the attacker can  simply resort to the negation of the output.

\begin{figurehere}
   \centering
    \includegraphics[scale=1,width=1.0\textwidth]{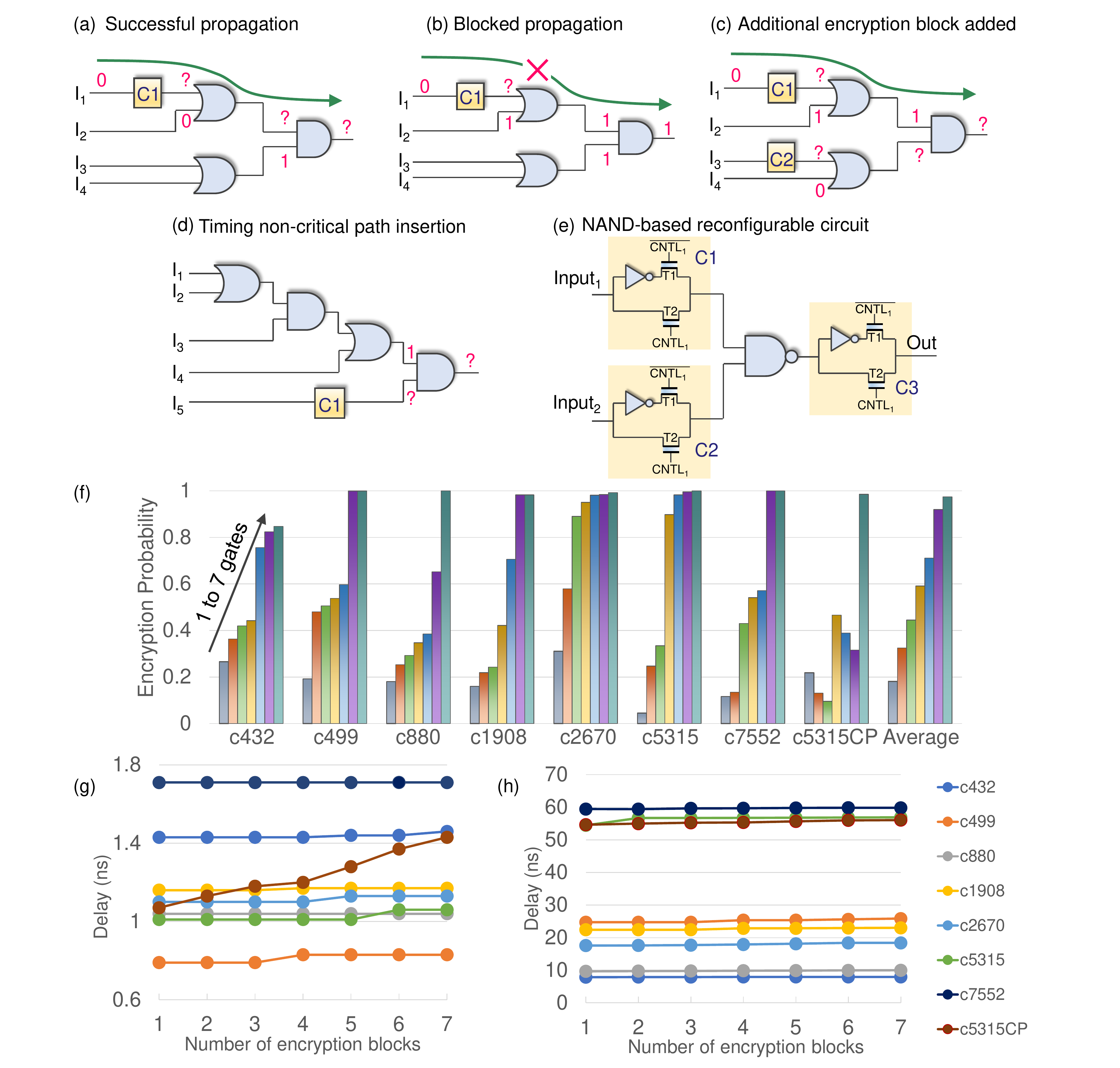}
        \caption{Analysis of placement  ultra-compact FeFET  encryption logic on  encryption probability, timing  and logic reconfigurability. (a) Successful propagation of an encrypted bit from C1. (b) Encrypted bit from C1 blocked by OR gate. (c) Enhancing encryption probability by inserting additional encryptor. (d) Placement of interconnect based encryption unit in a  noncritical timing path. (e) Reconfigurable logic based on NAND and active interconnect  based encryption logic. (f) Encryption probability with increasing number of encryption blocks on ISCAS85 benchmarks. (g) Critical path delay  with increasing number of encryption elements. (h) Sum of top100 critical path delays on ISCAS circuits.}
    \label{fig:figure4_propagation}
\end{figurehere}

Timing closure is one of the most critical challenges in ASIC/SoC designs with ever increasing clock rates\cite{vlsi'11, timingclosure'01, ICSICT'12}.
In this work, the aim is to incorporate encryption in the IC with minimal impacts on timing critical paths. Critical paths are the longest delay paths that limit clocking. Changing the standard logic gate design to adaptable camouflaging gate design\cite{IyengarG15} for security purposes increases the critical path delay. In the proposed method,  encryption gates are inserted in a non critical timing path to overcome the  potential timing failure. 
For example, in Fig.~\ref{fig:figure4_propagation}(d), the path from input I$_5$ to output is the least timing critical path, as it has the minimal number of gates from the input to output. In this case,  C1 is inserted in the logical branch from I$_5$ to the output. Though C1 is not in the timing critical path, it still logically  affects output as evident from Fig.~\ref{fig:figure4_propagation}(d).

In this study, ISCAS85 benchmarks\cite{ISCAS85} are simulated to analyze the encryption probability. Synopsis Design Compiler \cite{} is used  for logic synthesis. 
PRIMETIME\cite{primetime} is used for timing analysis. 
The simulations are based on the NCSU FreePDK 45 nm logic technology \cite{ncsu45} and a calibrated Verilog-A model of FeFET. Delay simulation of the encryption circuits is done with SPECTRE. 
Functional correctness is verified  with test vectors using Vivado Simulator \cite{vivado}. Random test vectors are generated for benchmarking. In this analysis, a non-critical path is chosen as a candidate for placement and the encryption circuit is placed  randomly in the chosen path. 
Then  test input pattern is applied to the modified  circuit with encryption blocks.
If the  output generates an incorrect result, then output is considered to be encrypted for this input pattern. In this work, the encryption probability is defined as the fraction of times we get the incorrect output out of the total number of attempted tests.

 Fig.~\ref{fig:figure4_propagation}(f) shows the number of encryption element versus encryption probability for ISCAS benchmarks(c432, c499, c880, c1908, c2670, c5315 and c7552).
C5315CP is the same benchmark as C5315, but with encryption blocks placed in a critical timing path and is used to show the  difference in impacts compared to the proposed method. The analysis shows 84.7\% to 100\% encryption probability  with an average 97.43\% with a total of 7 encryption blocks in the circuit. In general, the trend  shows that the encryption probability increases initially  with the increase in the number of encryption circuits.
However,  c5315CP does not show a monotonic increase in encryption probability with increase in the number of  encryption blocks. For c5315CP, as mentioned earlier, encryption gates are added in the same input to the output path.  This leads to double inversions in some cases, which in turn decreases the encryption probability.
Also, it is observed from Fig.~\ref{fig:figure4_propagation}(f) that a comparable level of encryption probability can be achieved with a smaller number of encryption blocks for many benchmarks. For example, in C499 increasing the number of encryption blocks from 6 to 7 does not change encryption probability and it maintains at 99.9\%. Similarly, also in C1908,  increasing the number of encryption blocks from 6 to 7 does not increase the  encryption probability  from 98.2\%. In C2670, increasing the number of encryption blocks from 5 to 6 increases the encryption probability only by 0.3\% from 98.1\% to 98.4\%.

Fig.~\ref{fig:figure4_propagation}(g) shows the critical timing path delay versus the added number of encryption circuits. The placement of 7 encryption blocks gives an average encryption probability of  97.43\% with the increase in most critical path delay by 2.24\% on average. It is observed that for most benchmarks adding encryption logic does not change the delay of the most critical path as gates are placed in different non-critical paths. For example, adding a single encryption gate  does not affect the critical path delay, and adding 6 gates changes the critical path delay by only 2.04\%.  Note that for  c5315CP,  the average delay on the most critical path gets worsened by 41.58\%  after the insertion of 7 gates. 
Further,  sum of delays on the top 100 critical paths is taken  for each of the  benchmarks to analyze the overall  delay impact in the IC after the placement of our encryption blocks. 
Insertion of multiple encryption units on the same path  can make a previously non-critical timing logic path to a critical timing path. Such occurrences  are restricted by  spreading out placement of encryption blocks  on different logic branches. Fig.~\ref{fig:figure4_propagation}(h) shows the sum of top 100 delays on ISCAS benchmarks. The impact of insertions is seen to be minimal. 
The placement of 5 encryption blocks gives an average encryption probability of 71.13\% with an overall delay increase by 1.34\%. The placement of 7 encryption blocks gives an average encryption probability of 97.43\% with an overall delay increase by 2.24\%.

Correlation of encryption probability with the placement of encryption logic is not linear. In the analysis, it is observed that encryption probability is at the highest if the encryption circuit is placed closer to the output node and becomes unpredictable as we move away from the output node due to logical masking.
The analysis of the placement of the  encryption logic in the same input-output path by varying the distance from output node is shown in supplementary materials Fig.~\ref{fig:replacement}.
The analysis on ISCAS85 benchmarks  demonstrates the ability to control output encryption probability without affecting the timing closure. It also shows,  addition of a large number of  encryption units is not   necessary to get a satisfactory encryption level, which is beneficial for overall area and power savings.  Typically,the placement of generic programmable gates \cite{JETCAS'14} worsen the timing closure challenges, where as the proposed techniques alleviate it by restricting the placements to non critical timing paths. Another advantage with our methodology is that instantiation of the interconnect based encryption gate from a standard cell logic library is possible on a need basis, making it easier for  automation and  eliminating  the  need  for  a  specialized  custom  design  of  cells.

The encryption analysis shows that random placement of encryption block in the logic circuit results in  different functional outputs. This concept can be extended to construct  reconfigurable logic gates by systematically placing the encryption logic around standard gates. Reconfigurable \label{Reconfigurable} logic is a known camouflaging method  to obfuscate the IP. Here the attacker will not be able to discern the logic and extract the correct netlist by observing the layout. Standard cell is the building block in ICs for logic operations. The more functions it has with the same layout, the harder to be attacked. The proposed active interconnect based encryption gate can be used in conjunction with standard cells to make a very easy implementable reconfigurable logic. 
An example is shown with NAND gate in Fig.~\ref{fig:figure4_propagation}(e).   Instantiating encryption logic in the inverter mode to the output of NAND gate makes the combination  an "AND" gate. Adding  encryption circuit to the inputs of NAND gates makes it further programmable. 
This makes the combination reconfigurable to NOR/OR. The internal programmed state of the instantiated interconnect based logic  and potential logic gates centered around NAND gate are shown in the  supplementary material Table.~\ref{fig:truth table of NAND}. 

 Adding the number of cells can improve the encyption stronger. 
 
 Adding randomlt dooes nt increase for example.

 Other is timing

 

 %

 In our work if we adopt a placement scheme such that it makes minimal impact in timing, yet get the circuit encrypted. 
 
 To test the effectiveness of the encryption circuit, we randomly choose a input-output path that is not on the critical path of the circuit and place the encryption circuit in the middle of the same path.  Then we used industry standard functional verification tools to measured correctness of circuit’s output on randomly generated input pattern. If the modified circuit output an incorrect
result, then the circuit is able to encrypt for this input pattern. Then we record the percentage of the circuit successfully encrypt the input.
 Fig. 6a shows the number of encryption element verses encryption probability for ISCAS benchmarks. We observes that the encryption probability would increase if the number of camouflage circuits increases in the circuit. Relation of encryption prob-ability and placement is more complicated. During the testing, we observes encryption probability is at the highest if the encryption circuit is placed on the output and at the lowest in the middle of input-output path as the logic masking effect form both input and output are the strongest.  We also tested the encryption logic placed on the same input-output path. For this placement strategy, encryption may be negated and result in a lower encryption probability. Although placement strategy is beyond the scope of this paper and needs further study, we demonstrated the ability to control output corrupt probability by varying number and placement of the camouflage circuit.

\section*{\large Peripheral Programming} \label{SecPeripheral_Programming}

The proposed peripheral circuitry using scan flipflops, as sketched in Fig.~\ref{fig:overview}(e), for reconfigurable encryption programming logic is shown in detail in Fig.~\ref{fig:peripheral_circuit}(b). Each of the FeFET pass transistors in the key logic needs to be programmed independently in the programming phase. Once programmed, circuits  operate in the logic mode of operation with logic mode voltages (i.e., FeFET state read voltages). Encryption logic is inserted along with the traditional logic circuitry as shown in Fig.~\ref{fig:peripheral_circuit}(b). The biasing voltage of the proposed scheme and the expected waveforms are given in Fig.~\ref{fig:peripheral_circuit}(a) and Fig.~\ref{fig:peripheral_circuit}(c) respectively. 
As discussed earlier,  complementary  programming states are required for the two FeFETs to store the encryption key. Once \textit{V}\textsubscript{TH} states are programmed into the corresponding FeFETs, a small read voltage is applied for logic operations. At first, the key sequence for programming is distributed to the  scan outputs of  flipflops ($S_{out}$, $\overline{S_{out}}$) through  the scan chain  using scan clocks ($S_{clk}$).

In the programming mode, depending on the key sequence, lower FeFET (F1) or upper FeFET(F2) get programmed to either HVT or LVT state. Scan outputs act as control signals to determine which FeFETs get to be HVT/LVT state. Typically in digital circuits, flipflop outputs (data outputs - $D_{out}/\overline{D_{out}}$, scan out - $S_{out}/\overline{S_{out}}$) are either  at a logic positive voltage or at a complementary GND voltage. FeFET requires a negative voltage for its writing process to be programmed as 
HVT \cite{S_Datta_reconfig}.
 This requirement requisites the scan output to be at a negative voltage at logic low for a straightforward implementation of control biasing scheme. However, it adds complexity in the flipflop design to make the scan output voltage to be biased at negative voltage at logic zero. To avoid substantial  flipflop modification by keeping scan logic zero at GND voltage and at the same time providing  required write  voltages to FeFETs, a  two step  programming process is proposed.

In the first step of programming,  all  FeFETs are made  HVT by providing negative write voltages at L1 and L2 and keep selector transistors (T1, T2) conducting. In the second step, only the required FeFETs are converted to  LVT. This is done by providing positive write voltages at L1 and L2 and  NMOS based selectors (T1, T2)  to be at ON/OFF state based on $S_{out}/\overline{S_{out}}$ .
For example if $S_{out}$ is high, it will make the T1 ON and T2 OFF, and positive write voltage will get transfered to the gate of F1(at \textit{int1} in Fig.~\ref{fig:peripheral_circuit}(b)) making the F1 LVT state. Since T2 is off, F2 will maintain the HVT state. In the logic mode, the L1 and L2 are provided with read voltages and selector transistors(T1 and T2) are turned ON by applying  $V_{read}$ at L4 at L5 as well. In summary, with the proposed method, some encryption blocks invert the input logic signal and some encryption blocks buffer the output  based on the provided encryption key sequence, such that overall functional obfuscation is achieved.  An alternative peripheral scheme  having one step write programming  with scan flipflop output being biased at negative voltage  for a logic zero is also given in the supplemental section (Fig.~\ref{fig:peripheral_circuit_v1}). 
 
\begin{figurehere}  
    \centering
    \includegraphics[scale=1.1,width=1\textwidth]{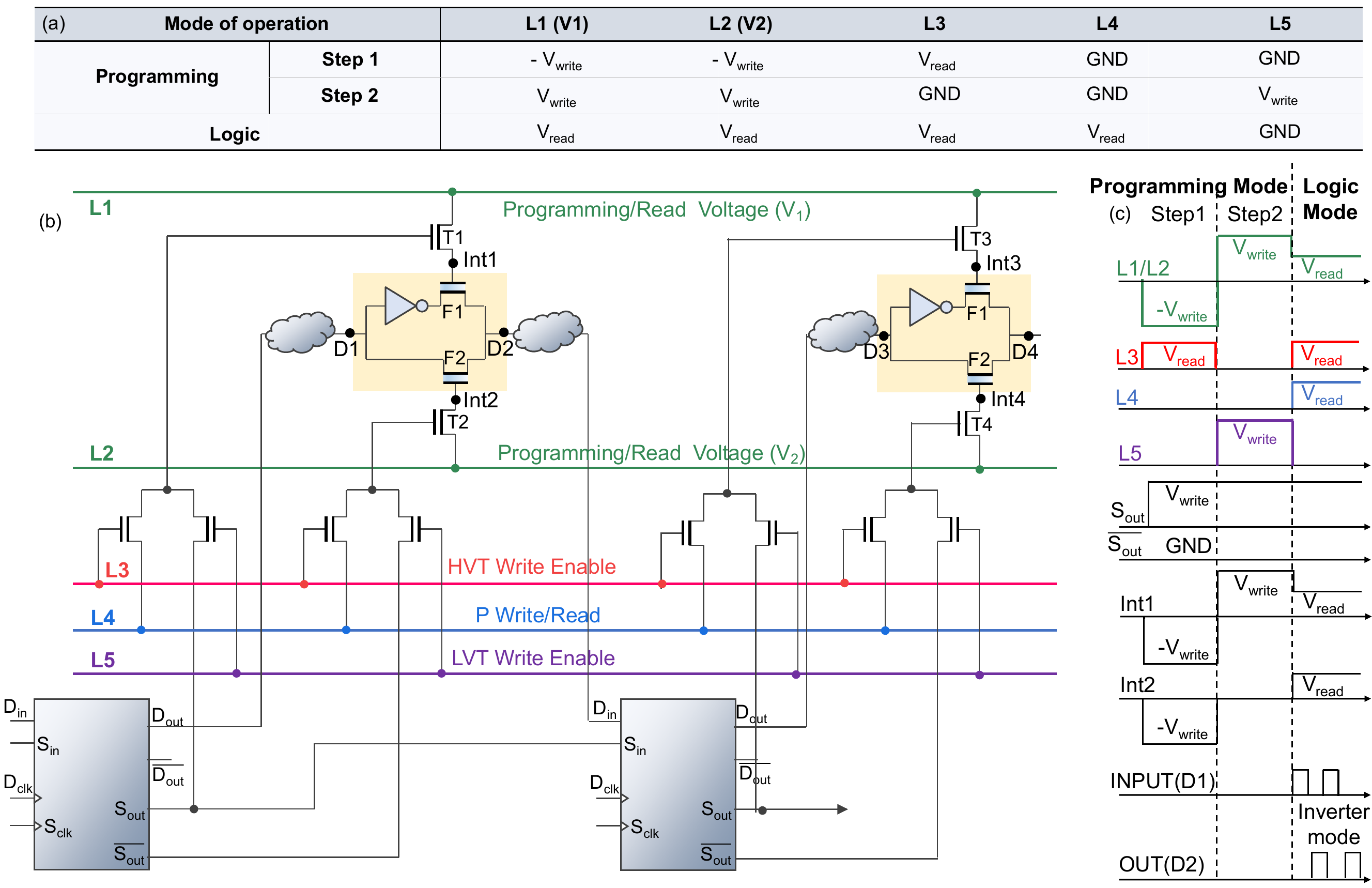}
    \caption{Circuit description of the proposed encryption key distribution utilizing  the existing data-scan flipflops in the IC. Grey bubbles indicates the  functional logic. The circuits in the yellow enclosure show encryption logic. (a) The biasing  scheme for the proposed peripheral scheme. (b) Schematic of the proposed encrypted IC with suggested peripheral circuitry by resusing the scan chain for the key distribution towards encryption blocks (c) Peripheral biasing waveforms in the programming and logic mode operation. Note, To minimize the routing lines and periphery required, most of the metal lines are shared temporally in the operations.}
    \label{fig:peripheral_circuit}
\end{figurehere}

\section*{\large Conclusion}

In this  work,  ultra compact active interconnect based  on ferroelectric FET  for hardware encryption is presented. FeFET leverages threshold voltage manipulation to attain run-time configurablity. The proposed   encryption circuit encompassing an inverter and an active interconnect, is layout obfuscated  and is capable  of producing either inverted or non-inverted output. This encryption circuit is fabricated and functionality is experimentally verified.  Further analysis showed that the placement of active interconnect encryption blocks in non critical timing  logic branches  produces satisfactory level of encryption without jeopardizing the timing closure requirement of ICs. Analysis on ISCAS benchmark shows a  97.43\%  encryption probability with an average delay increase of 3.68\% in the top 100 timing critical paths in ISCAS benchmarks. This work also introduced peripheral schemes for  programming the reconfigurable encryption keys by reusing the scan circuity and thereby eliminating the dedicated dynamic key input distribution logic and circuitry.

\section*{\large References}


\begin{thebibliography}{10}
\expandafter\ifx\csname url\endcsname\relax
  \def\url#1{\texttt{#1}}\fi
\expandafter\ifx\csname urlprefix\endcsname\relax\def\urlprefix{URL }\fi
\providecommand{\bibinfo}[2]{#2}
\providecommand{\eprint}[2][]{\url{#2}}

\bibitem{LogicLocking}
\bibinfo{author}{Rahman, M.~T.} \emph{et~al.}
\newblock \bibinfo{title}{Defense-in-depth: A recipe for logic locking to
  prevail}.
\newblock \emph{\bibinfo{journal}{Integr. VLSI J.}}
  \textbf{\bibinfo{volume}{72}}, \bibinfo{pages}{39–57}
  (\bibinfo{year}{2020}).
\newblock \urlprefix\url{https://doi.org/10.1016/j.vlsi.2019.12.007}.

\bibitem{Apple}
\bibinfo{author}{Anthony, S.}
\newblock \bibinfo{title}{iphone 5 a6 soc reverse engineered, reveals rare
  hand-made custom cpu, and tri-core gpu} (\bibinfo{year}{2012[online]}).
\newblock
  \urlprefix\url{https://www.extremetech.com/computing/136749-iphone-5-a6-soc-reverse-engineered-reveals-made-custom-cpu-and-a-tri-core-gpu}.

\bibitem{Mai'16}
\bibinfo{author}{{Erbagci}, B.}, \bibinfo{author}{{Erbagci}, C.},
  \bibinfo{author}{{Akkaya}, N. E.~C.} \& \bibinfo{author}{{Mai}, K.}
\newblock \bibinfo{title}{A secure camouflaged threshold voltage defined logic
  family}.
\newblock In \emph{\bibinfo{booktitle}{2016 IEEE International Symposium on
  Hardware Oriented Security and Trust (HOST)}}, \bibinfo{pages}{229--235}
  (\bibinfo{year}{2016}).

\bibitem{StateOfArt}
\bibinfo{author}{{Torrance}, R.} \& \bibinfo{author}{{James}, D.}
\newblock \bibinfo{title}{The state-of-the-art in semiconductor reverse
  engineering}.
\newblock In \emph{\bibinfo{booktitle}{2011 48th ACM/EDAC/IEEE Design
  Automation Conference (DAC)}}, \bibinfo{pages}{333--338}
  (\bibinfo{year}{2011}).

\bibitem{intel}
\bibinfo{author}{Hruska, J.~C.}
\newblock \bibinfo{title}{Intel's 14nm broadwell chip reverse engineered,
  reveals impressive finfets, 13-layer design} (\bibinfo{year}{2014}).
\newblock
  \urlprefix\url{https://www.extremetech.com/computing/193200-intels-14nm-broadwell-chip-reverse-engineered-reveals-impressive-finfets-13-layer-design}.

\bibitem{Rahman'14}
\bibinfo{author}{{Rahman}, M.~T.}, \bibinfo{author}{{Forte}, D.},
  \bibinfo{author}{{Shi}, Q.}, \bibinfo{author}{{Contreras}, G.~K.} \&
  \bibinfo{author}{{Tehranipoor}, M.}
\newblock \bibinfo{title}{Csst: Preventing distribution of unlicensed and
  rejected ics by untrusted foundry and assembly}.
\newblock In \emph{\bibinfo{booktitle}{2014 IEEE International Symposium on
  Defect and Fault Tolerance in VLSI and Nanotechnology Systems (DFT)}},
  \bibinfo{pages}{46--51} (\bibinfo{year}{2014}).

\bibitem{GDStoNetlist}
\bibinfo{author}{{Rajarathnam}, R.~S.}, \bibinfo{author}{{Lin}, Y.},
  \bibinfo{author}{{Jin}, Y.} \& \bibinfo{author}{{Pan}, D.~Z.}
\newblock \bibinfo{title}{Regds: A reverse engineering framework from gdsii to
  gate-level netlist}.
\newblock In \emph{\bibinfo{booktitle}{2020 IEEE International Symposium on
  Hardware Oriented Security and Trust (HOST)}}, \bibinfo{pages}{154--163}
  (\bibinfo{year}{2020}).

\bibitem{Randy'09}
\bibinfo{author}{Torrance, R.} \& \bibinfo{author}{James, D.}
\newblock \bibinfo{title}{The state-of-the-art in ic reverse engineering}.
\newblock In \bibinfo{editor}{Clavier, C.} \& \bibinfo{editor}{Gaj, K.} (eds.)
  \emph{\bibinfo{booktitle}{Cryptographic Hardware and Embedded Systems - CHES
  2009}}, \bibinfo{pages}{363--381} (\bibinfo{publisher}{Springer Berlin
  Heidelberg}, \bibinfo{address}{Berlin, Heidelberg}, \bibinfo{year}{2009}).

\bibitem{ISSCC'18}
\bibinfo{author}{{Akkaya}, N. E.~C.}, \bibinfo{author}{{Erbagci}, B.} \&
  \bibinfo{author}{{Mai}, K.}
\newblock \bibinfo{title}{A secure camouflaged logic family using
  post-manufacturing programming with a 3.6ghz adder prototype in 65nm cmos at
  1v nominal vdd}.
\newblock In \emph{\bibinfo{booktitle}{2018 IEEE International Solid - State
  Circuits Conference - (ISSCC)}}, \bibinfo{pages}{128--130}
  (\bibinfo{year}{2018}).

\bibitem{ETS'16}
\bibinfo{author}{{Nirmala}, I.~R.}, \bibinfo{author}{{Vontela}, D.},
  \bibinfo{author}{{Ghosh}, S.} \& \bibinfo{author}{{Iyengar}, A.}
\newblock \bibinfo{title}{A novel threshold voltage defined switch for circuit
  camouflaging}.
\newblock In \emph{\bibinfo{booktitle}{2016 21th IEEE European Test Symposium
  (ETS)}}, \bibinfo{pages}{1--2} (\bibinfo{year}{2016}).

\bibitem{HOST'18}
\bibinfo{author}{{Mohan}, P.}, \bibinfo{author}{{Akkaya}, N. E.~C.},
  \bibinfo{author}{{Erbagci}, B.} \& \bibinfo{author}{{Mai}, K.}
\newblock \bibinfo{title}{A compact energy-efficient pseudo-static camouflaged
  logic family}.
\newblock In \emph{\bibinfo{booktitle}{2018 IEEE International Symposium on
  Hardware Oriented Security and Trust (HOST)}}, \bibinfo{pages}{96--102}
  (\bibinfo{year}{2018}).

\bibitem{ISLPED'18}
\bibinfo{author}{Iyengar, A.~S.} \emph{et~al.}
\newblock \bibinfo{title}{Threshold defined camouflaged gates in 65nm
  technology for reverse engineering protection}.
\newblock In \emph{\bibinfo{booktitle}{Proceedings of the International
  Symposium on Low Power Electronics and Design}}, ISLPED '18
  (\bibinfo{publisher}{Association for Computing Machinery},
  \bibinfo{address}{New York, NY, USA}, \bibinfo{year}{2018}).
\newblock \urlprefix\url{https://doi.org/10.1145/3218603.3218641}.

\bibitem{VLSI'14}
\bibinfo{author}{{Roy}, K.}, \bibinfo{author}{{Sharad}, M.},
  \bibinfo{author}{{Fan}, D.} \& \bibinfo{author}{{Yogendra}, K.}
\newblock \bibinfo{title}{Computing with spin-transfer-torque devices:
  Prospects and perspectives}.
\newblock In \emph{\bibinfo{booktitle}{2014 IEEE Computer Society Annual
  Symposium on VLSI}}, \bibinfo{pages}{398--402} (\bibinfo{year}{2014}).

\bibitem{ESTCS'14}
\bibinfo{author}{{Sedighi}, B.}, \bibinfo{author}{{Hu}, X.~S.},
  \bibinfo{author}{{Nahas}, J.~J.} \& \bibinfo{author}{{Niemier}, M.}
\newblock \bibinfo{title}{Nontraditional computation using beyond-cmos
  tunneling devices}.
\newblock \emph{\bibinfo{journal}{IEEE Journal on Emerging and Selected Topics
  in Circuits and Systems}} \textbf{\bibinfo{volume}{4}},
  \bibinfo{pages}{438--449} (\bibinfo{year}{2014}).

\bibitem{ESTCS'15}
\bibinfo{author}{{Iyengar}, A.~S.}, \bibinfo{author}{{Ghosh}, S.} \&
  \bibinfo{author}{{Ramclam}, K.}
\newblock \bibinfo{title}{Domain wall magnets for embedded memory and hardware
  security}.
\newblock \emph{\bibinfo{journal}{IEEE Journal on Emerging and Selected Topics
  in Circuits and Systems}} \textbf{\bibinfo{volume}{5}},
  \bibinfo{pages}{40--50} (\bibinfo{year}{2015}).

\bibitem{2D-polarity}
\bibinfo{author}{Wu, P.}, \bibinfo{author}{Reis, D.}, \bibinfo{author}{Hu,
  X.~S.} \& \bibinfo{author}{Appenzeller, J.}
\newblock \bibinfo{title}{Two-dimensional transistors with reconfigurable
  polarities for secure circuits}.
\newblock \emph{\bibinfo{journal}{Nature Electronics}}
  \textbf{\bibinfo{volume}{4}}, \bibinfo{pages}{45--53} (\bibinfo{year}{2021}).
\newblock \urlprefix\url{https://doi.org/10.1038/s41928-020-00511-7}.

\bibitem{nature'20}
\bibinfo{author}{Lee, S.-J.} \emph{et~al.}
\newblock \bibinfo{title}{Programmable devices based on reversible solid-state
  doping of two-dimensional semiconductors with superionic silver iodide}.
\newblock \emph{\bibinfo{journal}{Nature Electronics}}
  \textbf{\bibinfo{volume}{3}}, \bibinfo{pages}{630–637}
  (\bibinfo{year}{2020}).
\newblock \urlprefix\url{https://doi.org/10.1038/s41928-020-00472-x}.

\bibitem{CCS'13}
\bibinfo{author}{Rajendran, J.}, \bibinfo{author}{Sam, M.},
  \bibinfo{author}{Sinanoglu, O.} \& \bibinfo{author}{Karri, R.}
\newblock \bibinfo{title}{Security analysis of integrated circuit
  camouflaging}.
\newblock In \emph{\bibinfo{booktitle}{Proceedings of the 2013 ACM SIGSAC
  Conference on Computer \&amp; Communications Security}}, CCS '13,
  \bibinfo{pages}{709–720} (\bibinfo{publisher}{Association for Computing
  Machinery}, \bibinfo{address}{New York, NY, USA}, \bibinfo{year}{2013}).
\newblock \urlprefix\url{https://doi.org/10.1145/2508859.2516656}.

\bibitem{S_Datta_reconfig}
\bibinfo{author}{{Dutta}, S.} \emph{et~al.}
\newblock \bibinfo{title}{Experimental demonstration of gate-level logic
  camouflaging and run-time reconfigurability using ferroelectric fet for
  hardware security}.
\newblock \emph{\bibinfo{journal}{IEEE Transactions on Electron Devices}}
  \textbf{\bibinfo{volume}{68}}, \bibinfo{pages}{516--522}
  (\bibinfo{year}{2021}).

\bibitem{greene201165nm}
\bibinfo{author}{Greene, J.} \emph{et~al.}
\newblock \bibinfo{title}{A 65nm flash-based fpga fabric optimized for low cost
  and power}.
\newblock In \emph{\bibinfo{booktitle}{Proceedings of the 19th ACM/SIGDA
  international symposium on Field programmable gate arrays}},
  \bibinfo{pages}{87--96} (\bibinfo{year}{2011}).

\bibitem{huang2014low}
\bibinfo{author}{Huang, K.}, \bibinfo{author}{Ha, Y.}, \bibinfo{author}{Zhao,
  R.}, \bibinfo{author}{Kumar, A.} \& \bibinfo{author}{Lian, Y.}
\newblock \bibinfo{title}{A low active leakage and high reliability phase
  change memory (pcm) based non-volatile fpga storage element}.
\newblock \emph{\bibinfo{journal}{IEEE Transactions on Circuits and Systems I:
  Regular Papers}} \textbf{\bibinfo{volume}{61}}, \bibinfo{pages}{2605--2613}
  (\bibinfo{year}{2014}).

\bibitem{zhao2009spin}
\bibinfo{author}{Zhao, W.}, \bibinfo{author}{Belhaire, E.},
  \bibinfo{author}{Chappert, C.} \& \bibinfo{author}{Mazoyer, P.}
\newblock \bibinfo{title}{Spin transfer torque (stt)-mram--based runtime
  reconfiguration fpga circuit}.
\newblock \emph{\bibinfo{journal}{ACM Transactions on Embedded Computing
  Systems (TECS)}} \textbf{\bibinfo{volume}{9}}, \bibinfo{pages}{1--16}
  (\bibinfo{year}{2009}).

\bibitem{tanachutiwat2010fpga}
\bibinfo{author}{Tanachutiwat, S.}, \bibinfo{author}{Liu, M.} \&
  \bibinfo{author}{Wang, W.}
\newblock \bibinfo{title}{Fpga based on integration of cmos and rram}.
\newblock \emph{\bibinfo{journal}{IEEE Transactions on Very Large Scale
  Integration (VLSI) Systems}} \textbf{\bibinfo{volume}{19}},
  \bibinfo{pages}{2023--2032} (\bibinfo{year}{2010}).

\bibitem{Ma'16}
\bibinfo{author}{George, S.} \emph{et~al.}
\newblock \bibinfo{title}{Nonvolatile memory design based on ferroelectric
  fets}.
\newblock In \emph{\bibinfo{booktitle}{2016 53nd ACM/EDAC/IEEE Design
  Automation Conference (DAC)}}, \bibinfo{pages}{1–6}
  (\bibinfo{publisher}{IEEE Press}, \bibinfo{year}{2016}).
\newblock \urlprefix\url{https://doi.org/10.1145/2897937.2898050}.

\bibitem{Xueqing'18}
\bibinfo{author}{{George}, S.} \emph{et~al.}
\newblock \bibinfo{title}{Symmetric 2-d-memory access to multidimensional
  data}.
\newblock \emph{\bibinfo{journal}{IEEE Transactions on Very Large Scale
  Integration (VLSI) Systems}} \textbf{\bibinfo{volume}{26}},
  \bibinfo{pages}{1040--1050} (\bibinfo{year}{2018}).

\bibitem{Lai'18}
\bibinfo{author}{{Li}, X.} \& \bibinfo{author}{{Lai}, L.}
\newblock \bibinfo{title}{Nonvolatile memory and computing using emerging
  ferroelectric transistors}.
\newblock In \emph{\bibinfo{booktitle}{2018 IEEE Computer Society Annual
  Symposium on VLSI (ISVLSI)}}, \bibinfo{pages}{750--755}
  (\bibinfo{year}{2018}).

\bibitem{ni2019ferro}
\bibinfo{author}{Ni, K.} \emph{et~al.}
\newblock \bibinfo{title}{Ferroelectric ternary content-addressable memory for
  one-shot learning}.
\newblock \emph{\bibinfo{journal}{Nature Electronics}}
  \textbf{\bibinfo{volume}{2}}, \bibinfo{pages}{521--529}
  (\bibinfo{year}{2019}).

\bibitem{8352114}
\bibinfo{author}{Ni, K.} \emph{et~al.}
\newblock \bibinfo{title}{Critical role of interlayer in hf0.5zr0.5o2
  ferroelectric fet nonvolatile memory performance}.
\newblock \emph{\bibinfo{journal}{IEEE Transactions on Electron Devices}}
  \textbf{\bibinfo{volume}{65}}, \bibinfo{pages}{2461--2469}
  (\bibinfo{year}{2018}).

\bibitem{9073266}
\bibinfo{author}{Thirumala, S.~K.}, \bibinfo{author}{Raha, A.},
  \bibinfo{author}{Narayanan, V.}, \bibinfo{author}{Raghunathan, V.} \&
  \bibinfo{author}{Gupta, S.~K.}
\newblock \bibinfo{title}{Non-volatile logic and memory based on reconfigurable
  ferroelectric transistors}.
\newblock In \emph{\bibinfo{booktitle}{2019 IEEE/ACM International Symposium on
  Nanoscale Architectures (NANOARCH)}}, \bibinfo{pages}{1--6}
  (\bibinfo{year}{2019}).

\bibitem{khan2020future}
\bibinfo{author}{Khan, A.~I.}, \bibinfo{author}{Keshavarzi, A.} \&
  \bibinfo{author}{Datta, S.}
\newblock \bibinfo{title}{The future of ferroelectric field-effect transistor
  technology}.
\newblock \emph{\bibinfo{journal}{Nature Electronics}}
  \textbf{\bibinfo{volume}{3}}, \bibinfo{pages}{588--597}
  (\bibinfo{year}{2020}).

\bibitem{mikolajick2020past}
\bibinfo{author}{Mikolajick, T.}, \bibinfo{author}{Schroeder, U.} \&
  \bibinfo{author}{Slesazeck, S.}
\newblock \bibinfo{title}{The past, the present, and the future of
  ferroelectric memories}.
\newblock \emph{\bibinfo{journal}{IEEE Transactions on Electron Devices}}
  \textbf{\bibinfo{volume}{67}}, \bibinfo{pages}{1434--1443}
  (\bibinfo{year}{2020}).

\bibitem{IEDM'16}
\bibinfo{author}{{Trentzsch}, M.} \emph{et~al.}
\newblock \bibinfo{title}{A 28nm hkmg super low power embedded nvm technology
  based on ferroelectric fets}.
\newblock In \emph{\bibinfo{booktitle}{2016 IEEE International Electron Devices
  Meeting (IEDM)}}, \bibinfo{pages}{11.5.1--11.5.4} (\bibinfo{year}{2016}).

\bibitem{IEDM'17}
\bibinfo{author}{{Dünkel}, S.} \emph{et~al.}
\newblock \bibinfo{title}{A fefet based super-low-power ultra-fast embedded nvm
  technology for 22nm fdsoi and beyond}.
\newblock In \emph{\bibinfo{booktitle}{2017 IEEE International Electron Devices
  Meeting (IEDM)}}, \bibinfo{pages}{19.7.1--19.7.4} (\bibinfo{year}{2017}).

\bibitem{IEDM'181}
\bibinfo{author}{{Jerry}, M.} \emph{et~al.}
\newblock \bibinfo{title}{Ferroelectric fet analog synapse for acceleration of
  deep neural network training}.
\newblock In \emph{\bibinfo{booktitle}{2017 IEEE International Electron Devices
  Meeting (IEDM)}}, \bibinfo{pages}{6.2.1--6.2.4} (\bibinfo{year}{2017}).

\bibitem{Jerry_2018}
\bibinfo{author}{Jerry, M.} \emph{et~al.}
\newblock \bibinfo{title}{A ferroelectric field effect transistor based
  synaptic weight cell}.
\newblock \emph{\bibinfo{journal}{Journal of Physics D: Applied Physics}}
  \textbf{\bibinfo{volume}{51}}, \bibinfo{pages}{434001}
  (\bibinfo{year}{2018}).
\newblock \urlprefix\url{https://doi.org/10.1088/1361-6463/aad6f8}.

\bibitem{Kai'18}
\bibinfo{author}{{Ni}, K.} \emph{et~al.}
\newblock \bibinfo{title}{Critical role of interlayer in hf0.5zr0.5o2
  ferroelectric fet nonvolatile memory performance}.
\newblock \emph{\bibinfo{journal}{IEEE Transactions on Electron Devices}}
  \textbf{\bibinfo{volume}{65}}, \bibinfo{pages}{2461--2469}
  (\bibinfo{year}{2018}).

\bibitem{5325880}
\bibinfo{author}{Chen, H.-T.}, \bibinfo{author}{Chang, C.-C.} \&
  \bibinfo{author}{Hwang, T.}
\newblock \bibinfo{title}{Reconfigurable eco cells for timing closure and ir
  drop minimization}.
\newblock \emph{\bibinfo{journal}{IEEE Transactions on Very Large Scale
  Integration (VLSI) Systems}} \textbf{\bibinfo{volume}{18}},
  \bibinfo{pages}{1686--1695} (\bibinfo{year}{2010}).

\bibitem{4359938}
\bibinfo{author}{Roy, J.~A.} \& \bibinfo{author}{Markov, I.~L.}
\newblock \bibinfo{title}{Eco-system: Embracing the change in placement}.
\newblock \emph{\bibinfo{journal}{IEEE Transactions on Computer-Aided Design of
  Integrated Circuits and Systems}} \textbf{\bibinfo{volume}{26}},
  \bibinfo{pages}{2173--2185} (\bibinfo{year}{2007}).

\bibitem{ibmz}
\bibinfo{title}{Ibm z}.
\newblock \urlprefix\url{https://www.ibm.com/it-infrastructure/z}.

\bibitem{trentzsch201628nm}
\bibinfo{author}{Trentzsch, M.} \emph{et~al.}
\newblock \bibinfo{title}{A 28nm hkmg super low power embedded nvm technology
  based on ferroelectric fets}.
\newblock In \emph{\bibinfo{booktitle}{2016 IEEE International Electron Devices
  Meeting (IEDM)}}, \bibinfo{pages}{11--5} (\bibinfo{organization}{IEEE},
  \bibinfo{year}{2016}).

\bibitem{mulaosmanovic2017switching}
\bibinfo{author}{Mulaosmanovic, H.} \emph{et~al.}
\newblock \bibinfo{title}{Switching kinetics in nanoscale hafnium oxide based
  ferroelectric field-effect transistors}.
\newblock \emph{\bibinfo{journal}{ACS applied materials \& interfaces}}
  \textbf{\bibinfo{volume}{9}}, \bibinfo{pages}{3792--3798}
  (\bibinfo{year}{2017}).

\bibitem{mulaosmanovic2020investigation}
\bibinfo{author}{Mulaosmanovic, H.} \emph{et~al.}
\newblock \bibinfo{title}{Investigation of accumulative switching in
  ferroelectric fets: enabling universal modeling of the switching behavior}.
\newblock \emph{\bibinfo{journal}{IEEE Transactions on Electron Devices}}
  \textbf{\bibinfo{volume}{67}}, \bibinfo{pages}{5804--5809}
  (\bibinfo{year}{2020}).

\bibitem{deng2020comprehensive}
\bibinfo{author}{Deng, S.} \emph{et~al.}
\newblock \bibinfo{title}{A comprehensive model for ferroelectric fet capturing
  the key behaviors: Scalability, variation, stochasticity, and accumulation}.
\newblock In \emph{\bibinfo{booktitle}{2020 IEEE Symposium on VLSI
  Technology}}, \bibinfo{pages}{1--2} (\bibinfo{organization}{IEEE},
  \bibinfo{year}{2020}).

\bibitem{freepdk45}
\bibinfo{title}{Freepdk45}.
\newblock \urlprefix\url{https://www.eda.ncsu.edu/wiki/FreePDK45:Contents}.

\bibitem{vlsi'11}
\bibinfo{author}{Kahng, A.~B.}, \bibinfo{author}{Lienig, J.},
  \bibinfo{author}{Markov, I.~L.} \& \bibinfo{author}{Hu, J.}
\newblock \emph{\bibinfo{title}{VLSI physical design: from graph partitioning
  to timing closure}} (\bibinfo{publisher}{Springer Science \& Business Media},
  \bibinfo{year}{2011}).

\bibitem{timingclosure'01}
\bibinfo{author}{Gosti, W.}, \bibinfo{author}{Khatri, S.} \&
  \bibinfo{author}{Sangiovanni-Vincentelli, A.}
\newblock \bibinfo{title}{Addressing the timing closure problem by integrating
  logic optimization and placement}.
\newblock In \emph{\bibinfo{booktitle}{IEEE/ACM International Conference on
  Computer Aided Design. ICCAD 2001. IEEE/ACM Digest of Technical Papers (Cat.
  No.01CH37281)}}, \bibinfo{pages}{224--231} (\bibinfo{year}{2001}).

\bibitem{ICSICT'12}
\bibinfo{author}{Yang, J.}, \bibinfo{author}{Shen, H.}, \bibinfo{author}{Liu,
  L.-k.} \& \bibinfo{author}{You, D.-s.}
\newblock \bibinfo{title}{Multi-mode timing closure of d6000 collective
  communication chip}.
\newblock In \emph{\bibinfo{booktitle}{2012 IEEE 11th International Conference
  on Solid-State and Integrated Circuit Technology}}, \bibinfo{pages}{1--3}
  (\bibinfo{year}{2012}).

\bibitem{IyengarG15}
\bibinfo{author}{Iyengar, A.} \& \bibinfo{author}{Ghosh, S.}
\newblock \bibinfo{title}{Threshold voltage-defined switches for programmable
  gates}.
\newblock \emph{\bibinfo{journal}{CoRR}}
  \textbf{\bibinfo{volume}{abs/1512.01581}} (\bibinfo{year}{2015}).
\newblock \urlprefix\url{http://arxiv.org/abs/1512.01581}.
\newblock \eprint{1512.01581}.

\bibitem{ISCAS85}
\bibinfo{author}{Hansen, M.}, \bibinfo{author}{Yalcin, H.} \&
  \bibinfo{author}{Hayes, J.}
\newblock \bibinfo{title}{Unveiling the iscas-85 benchmarks: a case study in
  reverse engineering}.
\newblock \emph{\bibinfo{journal}{IEEE Design Test of Computers}}
  \textbf{\bibinfo{volume}{16}}, \bibinfo{pages}{72--80}
  (\bibinfo{year}{1999}).

\bibitem{primetime}
\bibinfo{title}{Primetime from synopsys}.
\newblock
  \urlprefix\url{https://www.synopsys.com/support/training/signoff/primetime1-fcd.html}.

\bibitem{ncsu45}
\bibinfo{author}{Stine, J.~E.} \emph{et~al.}
\newblock \bibinfo{title}{Freepdk: An open-source variation-aware design kit}.
\newblock In \emph{\bibinfo{booktitle}{2007 IEEE International Conference on
  Microelectronic Systems Education (MSE'07)}}, \bibinfo{pages}{173--174}
  (\bibinfo{year}{2007}).

\bibitem{vivado}
\bibinfo{title}{Xilinx vivado design suite}.
\newblock
  \urlprefix\url{https://www.xilinx.com/products/design-tools/vivado.html}.

\bibitem{JETCAS'14}
\bibinfo{author}{Liu, B.} \& \bibinfo{author}{Wang, B.}
\newblock \bibinfo{title}{Reconfiguration-based vlsi design for security}.
\newblock \emph{\bibinfo{journal}{IEEE Journal on Emerging and Selected Topics
  in Circuits and Systems}} \textbf{\bibinfo{volume}{5}},
  \bibinfo{pages}{98--108} (\bibinfo{year}{2015}).

\end{thebibliography}

\section*{Acknowledgements}
This work is supported by the U.S. Department of Energy,
Office of Science, Office of Basic Energy Sciences Energy Frontier Research Centers program under Award Number DESC0021118.

\section*{Author contributions}

V.N., S.G., and K.N. proposed and supervised the project. Y.X. conducted the functionality verification simulation. T.Y. and Y.X. did the encryption analysis. T.Y. and N.J. did layout simulation. Y.K., S.D., and S.B. fabricated the FeFET devices, S.D. and Z.Z. performed the experimental characterizations of the FeFETs and encryption logic. All authors contributed to write up of the manuscript.

\section*{Competing interests}
The authors declare no competing interests.


\newpage

\renewcommand{\thefigure}{S\arabic{figure}}
\renewcommand{\thetable}{S\arabic{table}}
\setcounter{figure}{0}
\setcounter{table}{0}

\newpage
\centering
\title{\textbf{\Large Supplementary Materials}}
\begin{flushleft} 
\textbf{\large Device Fabrication}
\end{flushleft}
\justify
In this paper, the fabricated ferroelectric field effect transistor (FeFET) features a poly-crystalline Si/TiN (2 nm)/doped HfO\textsubscript{2} (8 nm)/SiO\textsubscript{2} (1 nm)/p-Si gate stack. The devices were fabricated using a 28 nm node gate-first high-K metal gate CMOS process on 300 mm silicon wafers. The ferroelectric gate stack process module starts with growth of a thin SiO\textsubscript{2} based interfacial layer, followed by the deposition of an 8 nm thick doped HfO\textsubscript{2}. 
A TiN metal gate electrode was deposited using physical vapor deposition (PVD), on top of which the poly-Si gate electrode is deposited. The source and drain n+ regions were obtained by phosphorous ion implantation, which were then activated by a rapid thermal annealing (RTA) at approximately 1000 $^\circ$C. This step also results in the formation of the ferroelectric orthorhombic phase within the doped HfO\textsubscript{2}.
For all the devices electrically characterized, they all have the same gate length and width dimensions of 1$\mu$m x 1$\mu$m, respectively.

\begin{flushleft} 
\textbf{\large Electrical Characterization}
\end{flushleft}
\justify
The FeFET device characterization was performed with
a Keithley 4200-SCS semiconductor parameter analyser. Two 4225-PMUs (pulse
measurement units) were utilized to make the pulsed current–voltage measurement.
In the experiment, program and erase pulses were applied and the pulsed \textit{I}\textsubscript{D}–\textit{V}\textsubscript{G}
(\textit{I}\textsubscript{D}, drain current; \textit{V}\textsubscript{G}, gate voltage) measurement was performed. The total sweep
duration is 5 ms. Note that, to minimize the charge trapping effects on the sensing of the programmed or erased state of the device, we inserted a delay of 100 ms between the measurement and the write pulses to allow a full trapped charge release. For the pulsed measurements, the current resolution is close to 3 nA in our set-up.
The reconfigurable block characterization was performed using two FeFETs on the same chip
and an externally connected inverter circuit (Texas Instruments 	
CD74AC04E). We connected the reconfigurable block with an inverter on a breadboard. Input pulses, FeFET memory write pulses, and evaluation pulses
were generated with an Keithley 4200-SCS. A 1.5V
amplitude VDD supply of the inverter was provided through an Agilent 81150A arbitrary function generator. The output voltage transient was sampled through an Tektronix TDS 2012B digital oscilloscope. All the write pulses have a pulse width of 10 $\mu$s. The input pulses have a pulse width of 100 $\mu$s and the evaluation pulse with a rising edge 5 $\mu$s ahead of the input rising edge and a falling edge 5 $\mu$s lagging behind the input falling edge. The large pulse width is chosen due to
the large parasitics in our set-up. In a fully integrated reconfigurable block, the operation speed will greatly improve, as shown in the single-FeFET measurement (successful write under 20 ns, ±4 V) in Fig.~\ref{fig:figure2_fefet}(d).

\begin{figure}[h]
    \centering
    \includegraphics[width=1\textwidth]{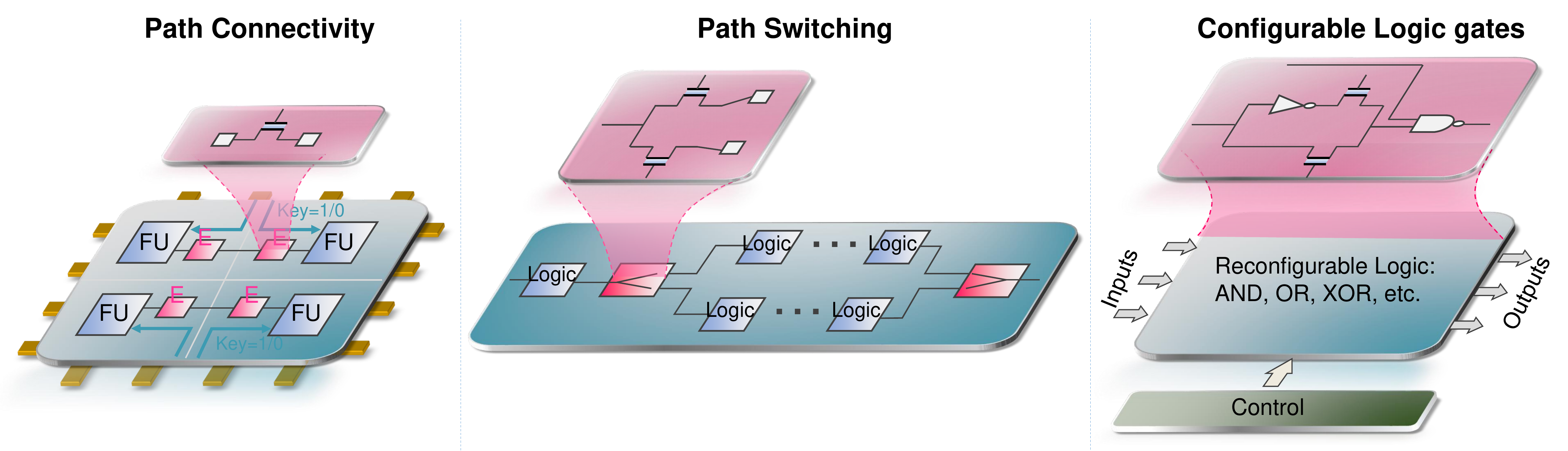}
    \caption{Other important applications that can benefit from ultra-compact FeFET active interconnect. It can be used as path connector, reconfigurable route switching, and reconfigurable logic. }
    \vspace{-3ex}
    \label{fig:other_applications}
\end{figure}

\newpage
\begin{flushleft} 
\textbf{\large Other Important Applications of FeFET Active Interconnect}
\end{flushleft}
\vspace{-7ex}
\justify
The variants of our ultra-compact FeFET active inter-connect design can be extended to apply in various chip design applications. Three potential applications that can be used in IC designs are listed in Fig.~\ref{fig:other_applications}. An example is the design of a configurable path connector capable of connecting/disconnecting  inputs to destination units. This is especially beneficial for controlling the logic signal flow towards redundant computation units. Inclusion of redundant functional units is a common method to develop reliable fault tolerant systems. In this application,  active ferroelectric based pass transistors can be utilized as path connectors and such units can be used to control the path connectivity between different functional units with ease.

Another potential application is configurable path switching which can essentially act as a router. Multiple active interconnect based pass transistors will be able route/block signals to different functional units as shown in Fig.~\ref{fig:other_applications}. 
In addition, active interconnect blocks can also be used to construct   reconfigurable logic gates by dynamically programming their control inputs.  Many combinations such as such as NAND, AND, OR, NOR, XOR, XNOR etc  are possible by the appropriate design (Fig.~\ref{fig:other_applications}).

\newpage
\begin{flushleft} 
\textbf{\large SPICE Simulations and Waveform Analysis}
\end{flushleft}

\begin{figure}
    \centering
    \includegraphics[ width=1.0\textwidth]{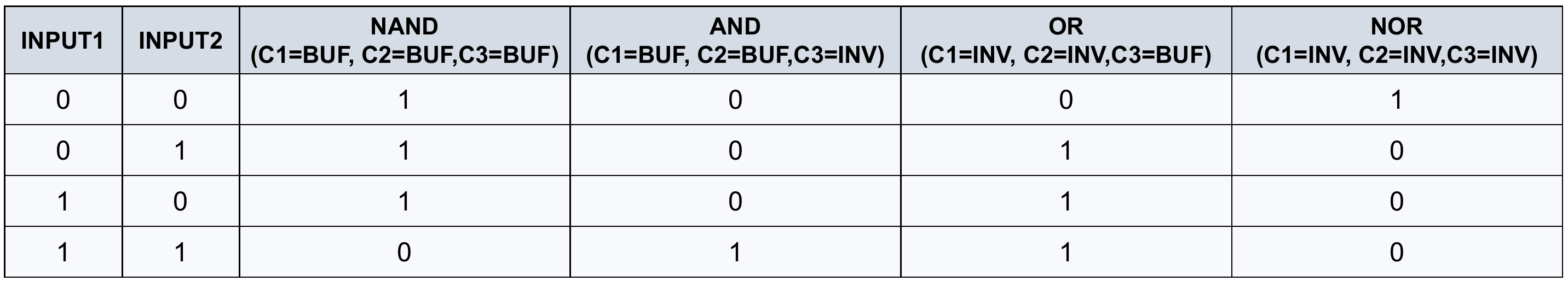}
    \captionof{table}{Truth table of NAND based reconfigurable logic}
    \label{fig:truth table of NAND}
\end{figure}

\begin{figure}[!ht]
    \centering
    \includegraphics[scale=1.0,width=0.5\textwidth]{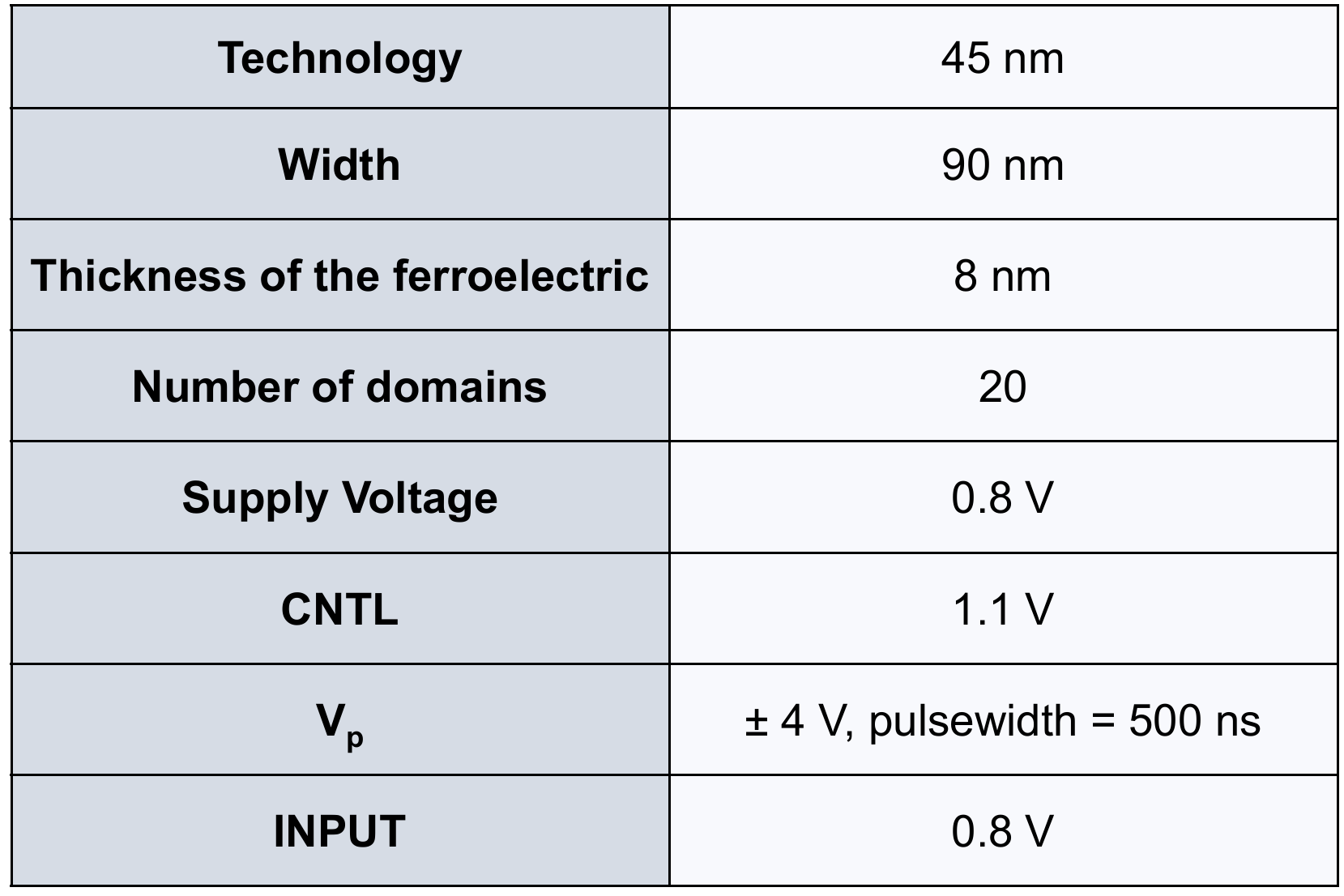}
    \captionof{table}{Simulation Parameters}
    \label{table:parameters_v1}
\end{figure}

\begin{figure}[!ht]
    \centering
    \includegraphics[scale=1.0,width=0.8\textwidth]{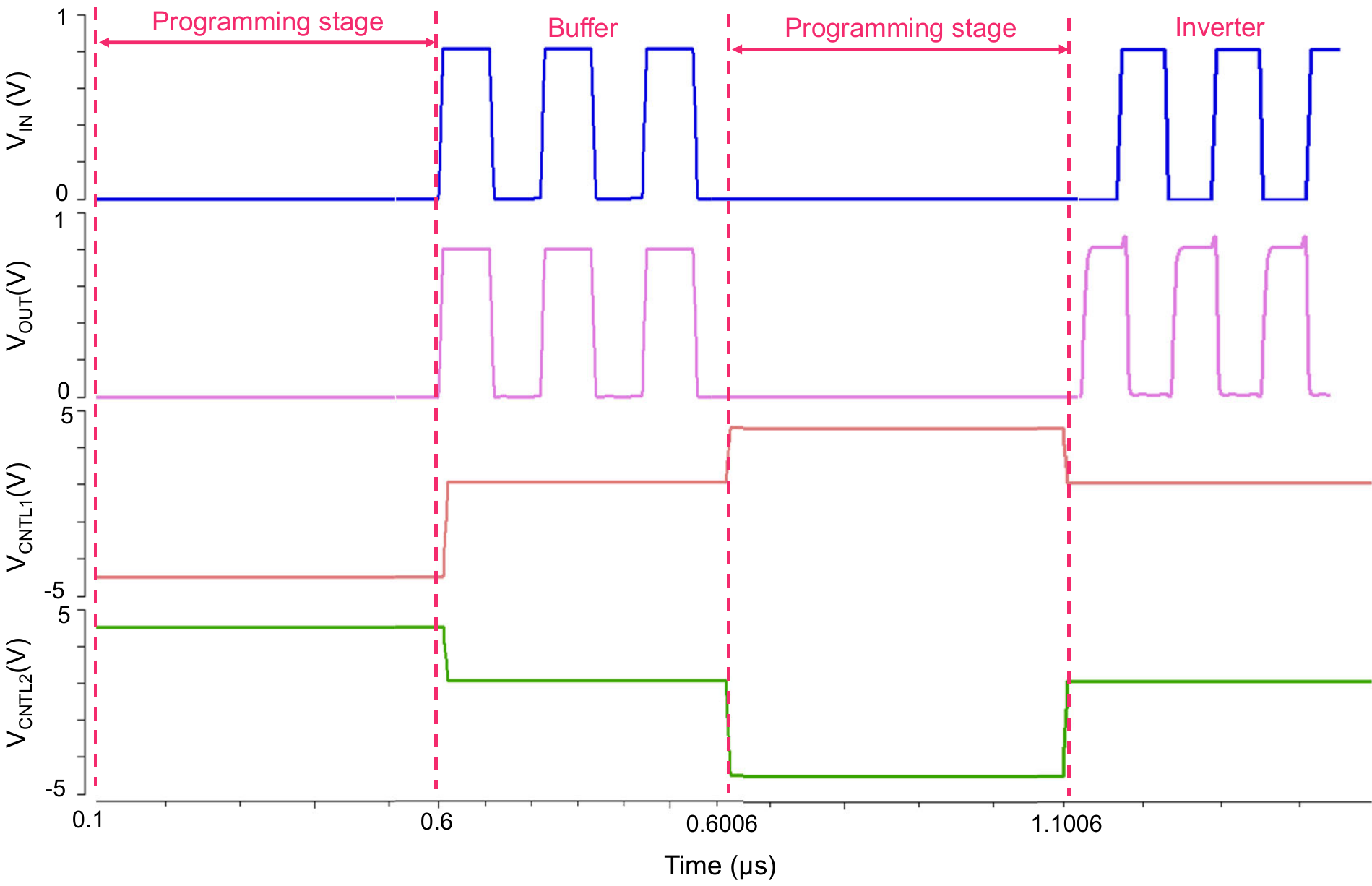}
    \caption{Simulated waveforms of the  dynamic programming of the FeFET active inter-connect encryption block (Fig.~\ref{fig:figure2_fefet}(e)).
    In the buffer mode programming stage, T1/ T2  is programmed to HVT/ LVT respectively by asserting  write voltage on CNTL signals. In the  logic mode,  read voltage is asserted on  CNTL signals and V$_{OUT}$ follows V$_{IN}$. In the inverter mode  programming stage,  T1/T2 is programmed to LVT/HVT respectively by asserting  write voltage on CNTL signals. In the logic mode,  read voltage is asserted on  CNTL signals and  V$_{OUT}$ shows inverted V$_{IN}$.  }
    \label{fig:dynamically programming}
\end{figure}

Dynamic programming simulation of the proposed active inter-connect based encryption block is shown in Fig.~\ref{fig:dynamically programming}. Spectre is used simulation verification and the schematic of the encryption unit is given in Fig.~\ref{fig:figure2_fefet}(e).  Table.~\ref{table:parameters_v1} shows simulation parameters.  Simulations are carried out using  NCSU FreePDK 45 nm technology\cite{ncsu45}. FeFET uses  verilog- A model to capture its characteristics. In this analysis,  a programming pulse (V$_P$ $\pm$4 V) with a pulsewidth of 500 ns is used to  set the threshold states  of two FeFET pass transistors. 

First, the block is programmed for buffer mode of operation. Second, the device is reprogrammed for inverter mode of operation. 
In the buffer mode of encryption, T1 (Fig.~\ref{fig:figure2_fefet}(f)) is programmed to HVT and T2 is programmed to LVT by asserting the write voltages in  CNTL  terminals. During this period, the polarizations of these two FeFETs are set in  opposite directions. In the evaluation mode, CNTL1 ($V_{CNTL1}$) and CNTL2 ($V_{CNTL2}$) are asserted with 1.1V and output (Out) follows the input (In).  This is shown as Programming stage/Buffer in Fig.~\ref{fig:dynamically programming}.

In the inverter mode of encryption, T1 (Fig.~\ref{fig:overview}(h)) is programmed to LVT and T2 is programmed to HVT by asserting the write voltages in  CNTL  terminals.
In the logic mode, CNTL1 ($V_{CNTL1}$) and CNTL2 ($V_{CNTL2}$) are asserted with 1.1V and logic input ($V_{IN}$) is set at 0.8 V for logic  high and 0 V for logic low. The output (Out)  shows the inverted input (In). In short,  the proposed encryption circuit can produce  two different outputs from the same input based on FeFETs' programmed states making a strong case for reverse engineering resilient hardware.

\newpage
\begin{flushleft} 
\textbf{\large Variation Analysis}
\end{flushleft}
\begin{figure}[h]
  \centering
  \includegraphics[scale=1.0, width=0.7\columnwidth]{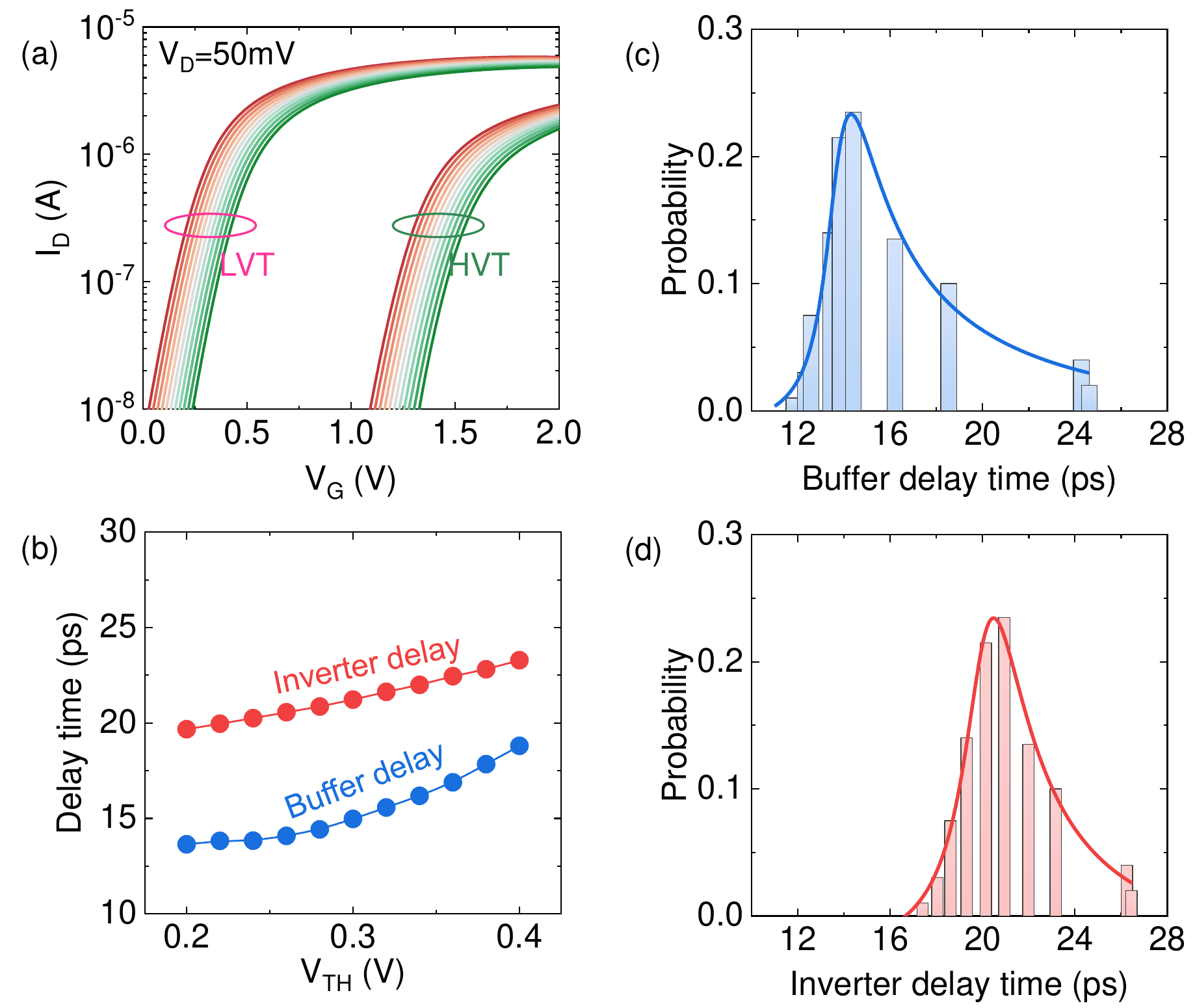}
  \caption{\label{Fig:figure3_variation_layout}Threshold voltage and delay variation analysis of the FeFET active inter-connect encryption  block. (a) Equivalent NMOS I-V characteristics of the FeFET in LVT and HVT states. (b) Delay of the proposed reconfigurable encryption circuit with equivalent nmos transistors in buffer mode and inverter mode. (c) Probability distribution of the delay in  reconfigurable encryption circuit with V$_T$ variation when working as a buffer. (d) Probability distribution of the delay in encryption circuit with V$_T$ variation when working as an inverte.}
\end{figure}

Variation analysis is conducted to study the delay impact of our proposed circuit.
Fig.~\ref{Fig:figure3_variation_layout}(a) shows I$_{d}$ - V$_{g}$ characteristics of the calibrated NMOS transistor of the equivalent FeFET in  LVT and HVT states. Monte Carlo simulations are performed to model threshold voltage  gaussian variation.  The delay variation  of the encryption circuits in inverter mode and buffer mode with respect to threshold voltage change are shown in Fig.~\ref{Fig:figure3_variation_layout}(b). The probabilistic  distribution of the buffer delay and inverter delay with respect to threshold voltage variation is given in Fig.~\ref{Fig:figure3_variation_layout}(c),(d).
The analysis shows an overall  delay variation of 6.4 ns/4.5 ns for the proposed circuit in  respective  buffer/inverter modes .
 

\newpage
\begin{flushleft} 
\textbf{\large Peripheral Circuits}
\end{flushleft}
\vspace{-4ex}

A peripheral scheme for encryption key distribution is introduced in  Fig.~\ref{fig:peripheral_circuit}. It uses a two step write process to eliminate the negative voltage requirement for logic zero scan output from the flipflop.  Here, an alternate design to   eliminate  two step write process is shown. However the process requires a flipflop where logic zero scan output is  biased at a negative voltage. Fig.~\ref{fig:peripheral_circuit_v1} shows the alternate circuit in programming mode and logic mode.  

\begin{figurehere}
    \centering
    \includegraphics[scale=1.0,width=1\textwidth]{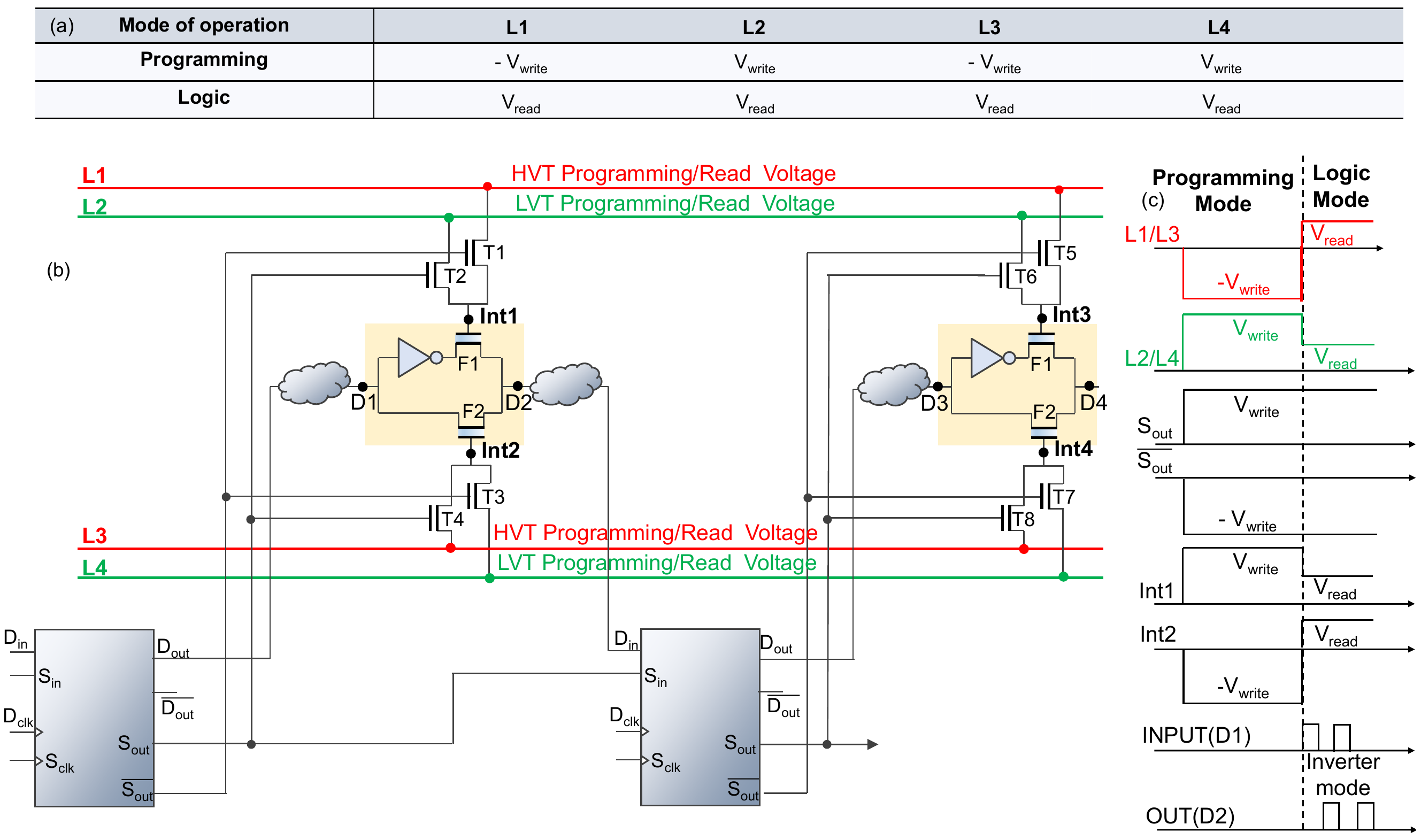}
    \caption{Circuit  description  of  the  proposed  encryption  key  distribution  with one step programming scheme. (a) The biasing for the peripheral scheme. (b) The peripheral circuit for programming. (c) Peripheral biasing waveforms in the programming and logic mode operation.}
    \label{fig:peripheral_circuit_v1}
\end{figurehere}

Programming mode  and logic mode operating voltages are exerted on  L1, L2, L3, L4 ( Fig.~\ref{fig:peripheral_circuit_v1}). 
In the programming  mode, the two selector transistors in the top (T1 \& T2)  and the two selector transistors in the bottom (T3 \& T4) act as switches to enable the transfer of  programming write  voltages to the encryption circuits. 
Complementary  write voltages are given to the upper and lower part of encryption unit.  $Scan\_out$ and $\overline{\mbox{$scan\_out$}}$ enable a pair of selector transistors  and  establishes a  connection to FeFETs (F1, F2)  either from  L1 \& L4 (buffer mode) or  from L2 \& L3(inverter mode).
With this version of peripheral circuit,  programming  of both the FeFETs in the encryption circuit (F1 \& F2)  to complementary $V_{T}$ state is done in one step by applying  corresponding  positive or negative write  voltages as shown in  Fig.~\ref{fig:peripheral_circuit_v1}(c). 
In logic mode,  read voltages are exerted  on L1, L2, L3 and L4 and  these voltages get  transferred to the gate of the ferroelectric FETs (F1, F2). Depending on the programmed state of the FeFETs the encryption block  will either produce  a buffered or an inverted version of input. The biasing for programming and logic modes are given in Fig.~\ref{fig:peripheral_circuit_v1}(a). Expected waveform on various metal lines during the operation is  given in Fig.~\ref{fig:peripheral_circuit_v1}(c).

\newpage
\begin{flushleft} 
\textbf{\large Benchmarking Analysis}
\end{flushleft}
\vspace{-4ex}

\begin{figure}[h]
    \centering
    \includegraphics[scale=1.0,width=1.0\textwidth]{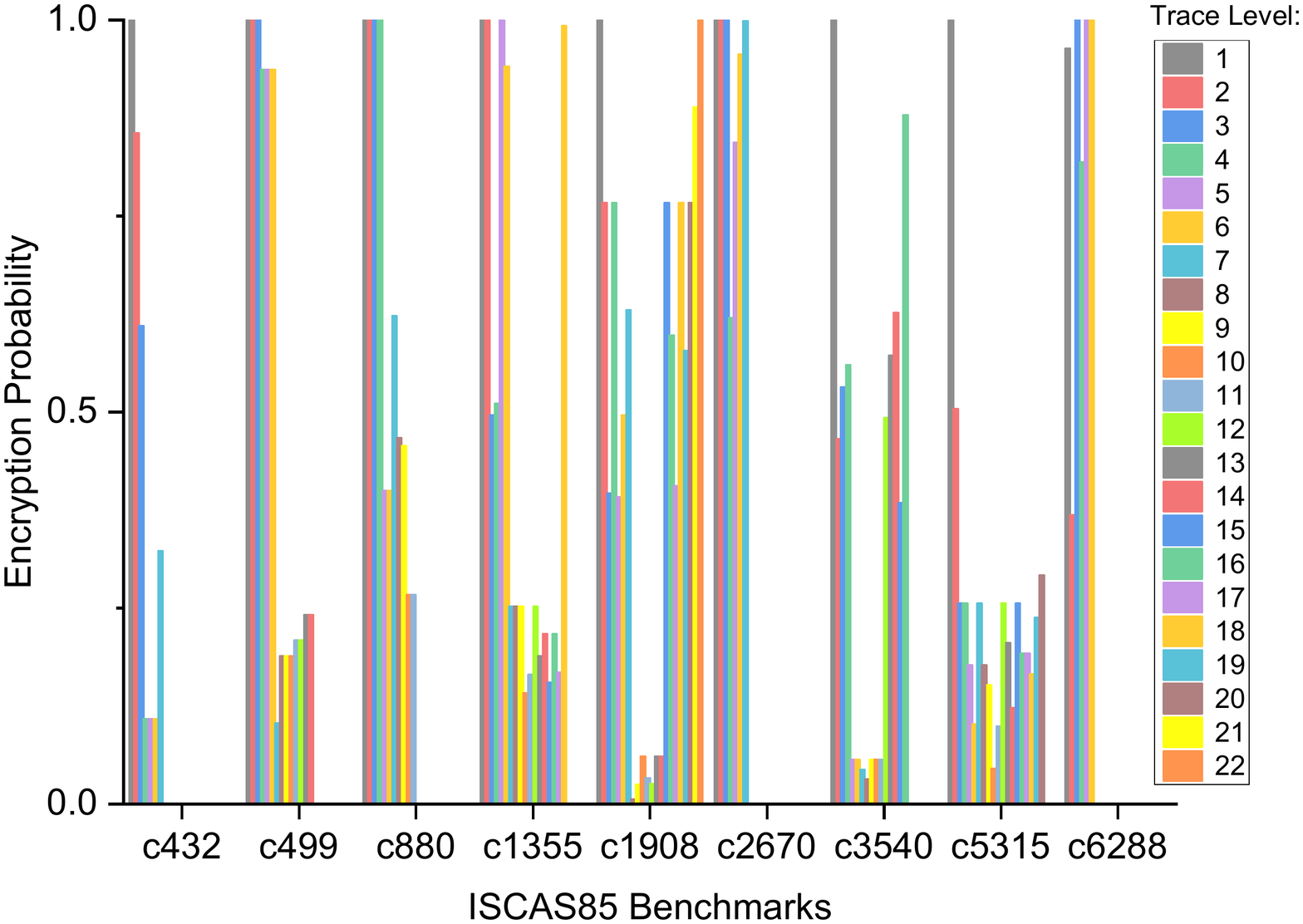}
    \caption{Impact on encryption probability when encryption unit is placed at increasing  logic distances from the output pin. }
    \label{fig:replacement}
\end{figure}

In this section,  the impact of the position of placement  of the proposed encryption block on encryption probability is analysed. 
Analysis begins  by  placing  one encryption block at the output of critical path and measuring the encryption probability.  Then the encryption block is moved to the input of the current gate and the corresponding encryption probability is recorded. This placement  process is repeated till the primary input of the critical path is reached. 
The experiment results with ISCAS85  benchmarks are shown in
 Fig.~\ref{fig:replacement}. The different levels denote the gate distance from the output. C432 and C499 are  small circuits with lesser than 20 levels from the output.
 
 Fig.~\ref{fig:replacement} indicates that for  most of the benchmark circuits, encryption probability is at the highest when at level 1. This is where the encryption circuit is placed closest to the output.  Then as the encryption element moves away to the center of critical path,  encryption probability decreases. This is attributed to the increased potential for logic masking effect  with the increase in distance from  the output. Also it is observed that, encryption probability further gets increased with further movement towards the input.As the placement moves closer to input, there is  an increased potential for higher fanout and more  logic branches getting influenced  by the encryption logic  and hence the potential for altering multiple outputs. In  benchmark C5315, encryption probability  decreases with  level 1, level2, level3 etc. A  similar behaviour is observed  with C432 as well. 
 C2670 and C1908 show increased encryption probability while moving encryption unit towards the input.

Note, for all  encryption and timing analysis in this article,  below motioned  methodology is adopted.  PRIMETIME\cite{primetime} is used for timing analysis. SPECTRE simulation  is used  to model the delay of our encryption circuit. Verilog-A is used  to capture FeFET behaviour. The simulations are based on NCSU FreePDK 45 nm  technology\cite{ncsu45}. Functional correctness  is verified with Xilinx Vivado  \cite{vivado} on the circuits with encryption blocks for the generated test vectors.In addition  these analyses incorporated a timing scaling factor to match the SPICE simulation delay with PRIME TIME library delay . The proposed circuit has used a  best case scaling factor of 0.94 and a worst case scaling factor of 2.29 to match the PRIMETIME library delay values.
\begin {comment}
}
\textcolor{green}{Does not know where this comes from. }\textcolor{yellow}{ modified the sentence a bit. there is a timing delay mismatch between what is measured from the spectre 45nm simulation and what is used in prime time library for the same gates. So we used a scaled value to match the rest of the gates in the library. The scaling factors correspond what we measure for the inverter in sprectre to what is given in PRIME time lib for the same. I guess we can move this details to benchmarking placement analysis in supplemental section ,} \textcolor{blue}{Please do. I think it is more a distraction than explanation here.}
\end{comment}

\newpage
\begin{flushleft} 
\textbf{\large Layout \& Device Analysis}
\end{flushleft}
\vspace{-4ex}

\begin{figure}[ht!]
  \centering
  \includegraphics[scale=0.8, width=0.8\columnwidth]{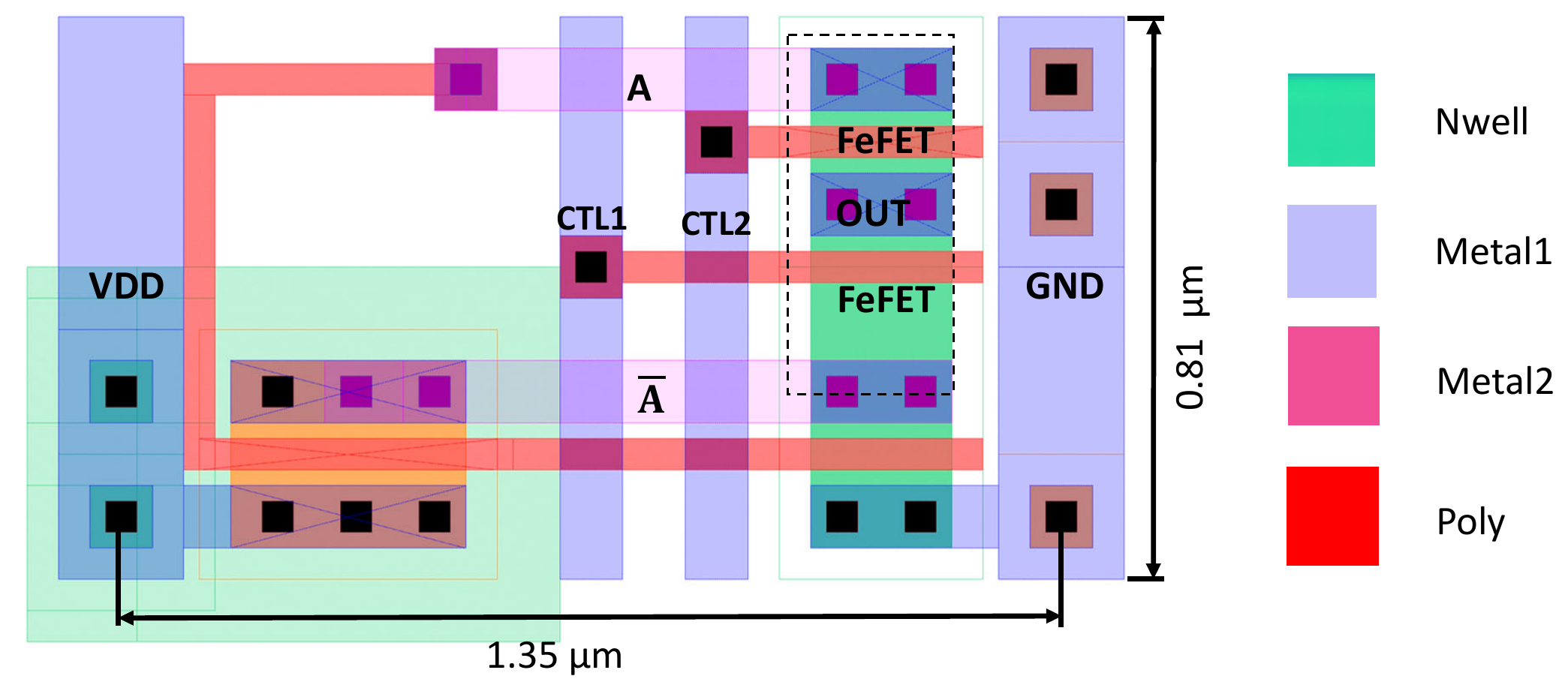}
  \caption{\label{fig:layout} Layout of a single active interconnect based encryption block.}
\end{figure}

Fig.~\ref{fig:layout} shows the layout of one encryption block consisting of two FeFETs and one inverter. The block has 1.35 $\mu$m (30F) width and 0.81 $\mu$m (18F) height. The area is calculated as 1.09 $\mu$m$^2$. Note, compared to the TVD implementation \cite{S_Datta_reconfig} needing  30 transistors, our implementation needs only 4 transistors while maintaining the camouflaging functionality.

\begin{figure}[ht]
    \centering
    \includegraphics[scale=1.0,width=0.8\textwidth]{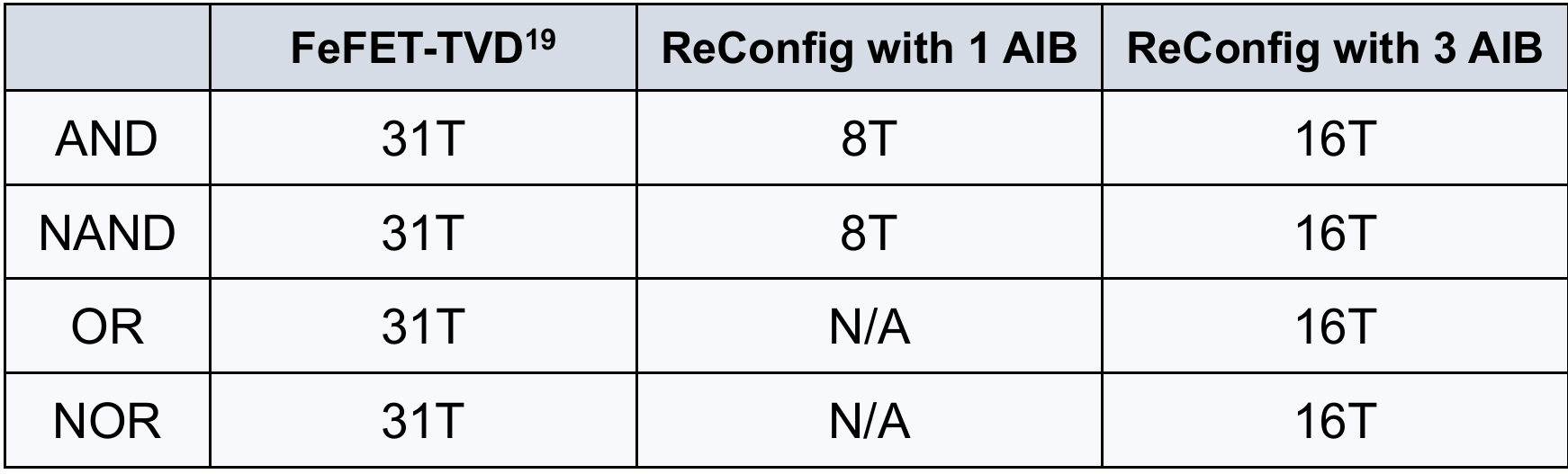}
    \captionof{table}{Number of transistor required for previous FeFET-TVD\cite{S_Datta_reconfig} Circuit and Proposed Reconfigurable logic with active interconnect blocks}
    \label{table:number of transistors}
\end{figure}

Table.~\ref{table:number of transistors} gives a comparison between  TVD implementation\cite{S_Datta_reconfig} and  reconfigurable logic  centred on NAND  gate (Tabe.~\ref{fig:truth table of NAND}) using active  interconnect blocks (AIB) based on the number of transistors.  Using one active interconnect block at the output of NAND gives reconfigurable AND/NAND logic. The above mentioned circuit takes 8 transistors. In addition, adding 2 more AIBs at the input of NAND give 4 reconfigurable logic possibilities with 16 Transistors. 

\end{document}
